\documentclass[a4paper,11pt]{article}
\pdfoutput=1 %

\usepackage{styles/jheppub} %

\usepackage[T1]{fontenc} %

\usepackage{styles/main}
\usepackage{styles/defs}

\graphicspath{{assets/}{assets/plots/res/}{assets/plots/res-tmp/}}

\title{\eko: Evolution Kernel Operators}

\collaborationImg{\includegraphics[width=.3\textwidth]{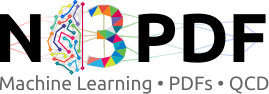}}

\author[a]{Alessandro Candido,}
\author[a,1]{Felix Hekhorn,\note{Corresponding author.}}
\author[b,c]{Giacomo Magni}

\affiliation[a]{Tif Lab, Dipartimento di Fisica, Universit\`a di Milano and\\
INFN, Sezione di Milano, Via Celoria 16, I-20133 Milano, Italy}
\affiliation[b]{Department of Physics and Astronomy, Vrije Universiteit, NL-1081 HV Amsterdam}
\affiliation[c]{Nikhef Theory Group, Science Park 105, 1098 XG Amsterdam, The Netherlands}

\emailAdd{alessandro.candido@mi.infn.it}
\emailAdd{felix.hekhorn@mi.infn.it}
\emailAdd{gmagni@nikhef.nl}

\preprint{TIF-UNIMI-2022-2, Nikhef-2022-003}

\abstract{
    We present a new \qcd{} evolution library for unpolarized parton
    distribution functions: \eko{}.
    The program solves \dglap{} equations up to next-to-next-to-leading order.
    The unique feature of \eko{} is the computation of solution operators,
    which are independent of the boundary condition, can be stored
    and quickly applied to evolve several initial \pdf{}s.
    The \eko{} approach combines the power of $N$-space solutions with the
    flexibility of a $x$-space delivery, that allows for an easy
    interface with existing codes.
    The code is fully open source and written in Python, with a modular
    structure in order to facilitate usage, readability and possible
    extensions.
    We provide a set of benchmarks with similar available tools, finding good
    agreement.
}

\begin{document} 
\maketitle
\flushbottom

\section{Introduction}
\label{sec:intro}
As we are entering the era of high-energy precision physics, theorists strive
to keep up with the experimental precision~\cite{Gao:2017yyd}.
The determination of parton distribution functions (\pdf{}s) is becoming
a major limiting factor and theory groups come up with more and more
involved procedures to improve the extraction~\cite{NNPDF:2017mvq,Hou:2019efy,Bailey:2020ooq}
eventually aiming for a one percent accuracy~\cite{NNPDF:2021njg}.
In order to achieve this goal a thorough treatment of theoretical
uncertainties is required~\cite{AbdulKhalek:2019ihb},
that so far was challenging with the current
state-of-the-art codes.
In this paper, we present \eko{} a new \qcd{} evolution library that matches
the requirements and desiderata of this new era.

\eko{} solves the evolution
equations~\cite{Altarelli:1977zs,Gribov:1972ri,Dokshitzer:1977sg} in Mellin
space (see \cref{sec:theory:mellin}) to allow for simpler solution algorithms
(both iterative and approximated).
Yet, it provides result in momentum fraction space (see
\cref{sec:theory:interpolation}) to allow an easy interface with existing
codes.

\eko{} computes evolution kernel operators (EKO) which are independent of
the initial boundary conditions but only depend on the given theory settings.
The operators can thus be computed once, stored on disk and then reused in the
actual application. This method can lead to a significant speed-up when \pdf{}s
are repeatedly being evolved, as it is customary in \pdf{} fits.
This approach has been introduced by \fk{}~\cite{Ball:2008by,Ball:2010de,DelDebbio:2007ee}
and it is further developed here.

\eko{} is open-source, allowing easy interaction with users
and developers.
The project comes with a clean, modular, and maintainable codebase that guarantees
easy inspection and ensures it is future-proof.
We provide both a user and a developer documentation. So, not only a user
manual, but even the internal documentation is published, with an effort to make
it as clear as possible.

\eko{} currently implements the leading order (\lo{}),
next-to-leading order (\nlo{}) and next-to-next-to-leading order (\nnlo{})
solutions~\cite{Vogt:2004mw,Moch:2004pa,Blumlein:2021enk}.
However, it is organized in such a way that the addition of higher order corrections,
such as the next-to-next-to-next-to-leading order (\nnnlo{})~\cite{Moch:2021qrk},
can be achieved with relative ease.
This accuracy is needed to match the precision in the determination of the
matrix elements for several processes at LHC (see e.g.\ \cite{Duhr:2021vwj} and
references therein).
We also expose the associated variations of the various scales.

\eko{} correctly treats intrinsic heavy quark distributions,
required for studies of the heavy quark content of the nucleon~\cite{Ball:2022qks}.
While the treatment of intrinsic distributions in the evolution equations is
mathematically simple, as they decouple in a specific basis, their integration
into the full solution, including matching conditions, is non-trivial.
We also implement backward evolution, again including the non-trivial
handling of matching conditions.

\eko{} is another corner stone in a suite of tools that aims to
provide new handles to the theory predictions in the \pdf{} fitting
procedure. To obtain the theory predictions in a typical fitting procedure,
first, one needs to evolve a candidate \pdf{} up to the process scale,
and then convolute it with the respective coefficient function.
The process specific coefficient function can be stored in the
\pineappl{} format~\cite{Carrazza_2020,christopher_schwan_2022_5846421}.
\eko{} is interfaced with \pineappl{} to produce
interpolation grids that can be directly used in a \pdf{} fit,
avoiding to do the evolution on the fly, but beforehand.
\eko{} is also powering \yadism{}~\cite{yadism} a new \dis{}
coefficient function library.

\eko{} adopts Python as a programming language opting for a high-level language
which is easy to understand for newcomers.
In particular, with the advent of Data Science and Machine Learning, Python has
become the language of choice for many scientific applications, mainly driven
by the large availability of packages and frameworks.
We decided to write a code that can be used by everyone who needs \qcd{}
evolution, and to make it possible for applications that are not supported yet
to be built on top of the provided tools and ingredients.
For this reason the code is developed mainly as a library, that contains
physics, math, and algorithmic tools, such as those needed for managing or
storing the computed operators. As an example we also expose a runner, making
use of the built library to deliver an actual evolution application. 
We apply modern best practices for software development, such as automated
tests, Continuous Integration (CI), and Continuous Deployment (CD), to ensure a
high quality of coding standard and a routinely checked code basis.

\paragraph{References} The open-source repository is available at:
\begin{center}
\url{https://github.com/N3PDF/eko}
\end{center}
In the following we do not attempt to give a complete overview over all
provided features and options, but limit ourselves to a brief review. The
full documentation instead can be accessed at:
\begin{center}
\url{https://eko.readthedocs.io/en/latest/}
\end{center}
This document is also regularly updated and extended upon the
implementation of new features.

\section{Theory Overview}
\label{sec:theory}
We do not attempt to give a full review of the underlying theory
here as it is known since a long time and discussed extensive elsewhere
(see e.g.\ \cite{Peskin:1995ev,Ellis:1996mzs} and references therein).
We refer the interested reader to the specific references given in the following and to
the accompanying online documentation where instead we give a detailed
overview. All sections in the following have an equivalent section in
the online documentation. Also the respective code implementations of the
various ingredients contain relevant information and are also accessible
in the documentation via the API section.

The central equations that \eko{} is solving are the
Dokshitzer-Gribov-Lipatov-Altarelli-Parisi (\dglap) evolution
equations~\cite{Altarelli:1977zs,Gribov:1972ri,Dokshitzer:1977sg} given by
\begin{equation}
    \muF^2 \dv{\vb f}{\muF^2}{}(x,\muF^2) = \vb P (a_s(\muR^2),\muF^2) \otimes \vb f(\muF^2)
    \label{eq:dglap}
\end{equation}
where $\vb f(x,\muF^2)$ is a vector of \pdf{}s over flavor space with $x$ the
momentum fraction and $\muF^2$ the factorization scale.
The main ingredients to \cref{eq:dglap} are the Altarelli-Parisi splitting
functions $\vb P(a_s(\muR^2),x,\muF^2)$~\cite{Moch:2004pa,Vogt:2004mw}, which
are matrices over the flavor space.
Finally, $\otimes$ denotes the multiplicative (or Mellin) convolution.

The splitting functions $\vb P(a_s(\muR^2),x,\muF^2)$ expose a perturbative
expansion in the strong coupling $a_s(\mu_R^2)$:
\begin{equation}
    \vb P(a_s(\muR^2),x,\muF^2) = a_s(\muR^2) \vb P^{(0)}(x,\muF^2)
    + \qty[a_s(\muR^2)]^2 \vb P^{(1)}(x,\muF^2)
    + \qty[a_s(\muR^2)]^3 \vb P^{(2)}(x,\muF^2)
    + \ldots
\end{equation}
which is currently known at \nnlo{}~\cite{Moch:2004pa,Vogt:2004mw,Blumlein:2021enk} and is under
investigation for \nnnlo{}~\cite{Moch:2021qrk}.
In a first step, the renormalization scale $\muR$ and the factorization scale
$\muF$ can be assumed to be equal $\muR = \muF$ and the renormalization scale
dependence can be restored later on. The variation of the ratio $\muR/\muF$ can
be considered as an estimated to missing higher order
uncertainties (\mhou{})~\cite{AbdulKhalek:2019ihb}.

In order to solve \cref{eq:dglap} a series of steps has to be taken, and we
highlight these steps in the following sections.

\subsection{Mellin space}
\label{sec:theory:mellin}
The presence of the derivative on the left-hand-side and the convolution on the
right-hand-side turns \cref{eq:dglap} into a set of coupled
integro-differential equations which are non-trivial to solve.

A possible strategy in solving \cref{eq:dglap} is by tackling the problem
head-on and iteratively solve the integrals and the derivative by taking small
steps: we refer to this as \enquote{$x$-space solution}, as the solution uses
directly momentum space and this approach is adopted, e.g., by \apfel{}~\cite{Bertone:2013vaa},
\hoppet{}~\cite{Salam:2008qg}, and \qcdnum{}~\cite{Botje:2010ay}.
However, this approach becomes quite cumbersome when dealing with higher-order
corrections, as the solutions becomes more and more involved.

We follow a different strategy and apply the Mellin transformation $\Md$
\begin{equation}
    \tilde g(N) = \Md\qty[g(x)](N) = \int\limits_0^1\!\dd{x} x^{N-1} g(x)
\end{equation}
where, as well here as in the following, we denote objects in Mellin space by a
tilde.
This approach is also adopted by \pegasus{}~\cite{Vogt:2004ns} and \fk{}~\cite{Ball:2008by,Ball:2010de,DelDebbio:2007ee}.
The numerically challenging step is then shifted to the treatment of the Mellin
inverse $\Md^{-1}$, as we eventually seek for results in $x$-space (see
\cref{sec:theory:interpolation}).

\subsection{Interpolation}
\label{sec:theory:interpolation}
Mellin space has the theoretical advantage that the analytical solution of the
equations becomes simpler, but the practical disadvantage that it requires
\pdf{}s in Mellin space. 
This constraint is in practice a serious limitation since most matrix element
generators~\cite{Buckley:2011ms} as well as the various generated coefficient
function grids (e.g.\ \pineappl{}~\cite{Carrazza_2020,christopher_schwan_2022_5846421},
\appl{}~\cite{Carli:2010rw} and \fastnlo{}~\cite{Britzger:2012bs}) are not
using Mellin space, but rather $x$-space.

This is bypassed in \pegasus{} by parametrizing the initial boundary condition
with up to six parameters in terms of the Euler beta function.
However, this is not sufficiently flexible to accommodate more complex analytic
forms, or even parametrizations in form of neural networks.

We are bypassing this limitation by introducing a Lagrange-interpolation~\cite{LagrangeInterpol,suli2003introduction} of the
\pdf{}s in $x$-space on arbitrarily user-chosen grids $\mathbb G$:
\begin{equation}
    f(x) \sim \bar f(x) = \sum_{j} f(x_j) p_j(x),  \quad \text{with}\,x_j\in \mathbb G
\end{equation}
For the usage inside the library we do an analytic Mellin transformation of the polynomials $\tilde p_j(N) = \Md\qty[p_j(x)](N)$.
For the interpolation polynomials $p_j$ we are choosing a subset with $N_{degree} + 1$ points of the interpolation grid $\mathbb G$
to avoid Runge's phenomenon~\cite{zbMATH02662492,suli2003introduction} and to avoid large cancellation in the Mellin transform.

\subsection{Strong coupling}
\label{sec:theory:coupling}
The evolution of the strong coupling $a_s(\muR^2) = \alpha_s(\muR^2)/(4\pi)$
is given by its renormalization group equation (\rge):
\begin{equation}
    \beta(a_s) = \muR^2\dv{a_s(\muR^2)}{\muR^2}{} = - \sum\limits_{n=0} \beta_n \qty[a_s(\muR^2)]^{2+n}
\end{equation}
and is currently known at 5-loop ($\beta_4$)
accuracy~\cite{Herzog:2017ohr,Luthe:2016ima,Baikov:2016tgj,Chetyrkin:2017bjc,Luthe:2017ttg}.

This is crucial for \dglap{} solution, indeed, since the strong coupling $a_s$
is a monotonic function of the renormalization scale in the perturbative
regime, we can actually consider a transformation of
\cref{eq:dglap}
\begin{equation}
    \dv{\vb{\tilde f}}{a_s}{}(N,a_s) = - \frac{\bm{\gamma}(N,a_s)}{\beta(a_s)} \vb {\tilde f}(N, a_s)
    \label{eq:dglap2}
\end{equation}
with $\bm{\gamma} = - \vb{\tilde P}$ the anomalous dimension and $\beta(a_s)$
the \qcd{} beta function, where the multiplicative convolution is reduced to an
ordinary product.

\subsection{Flavor space}
\label{sec:theory:flavor}
Next, we address the separation in flavor space: formally we can define the
flavor space $\Fd$ as the linear span over all partons (which we consider to be
the canonical one):
\begin{equation}
    \Fd = \Fd_{fl} = \vspan\qty(g, u, \bar u, d, \bar d, s, \bar s, c, \bar c, b, \bar b, t, \bar t)
\end{equation}

The splitting functions $\vb P$ become block-diagonal in the \enquote{Evolution
Basis}, a suitable decomposition of the flavor space: the singlet sector $\vb
P_S$ remains the only coupled sector over $\qty{\Sigma, g}$, while the full
valence combination $P_{ns,v}$ decouples completely (i.e.\ it is only coupling
to $V$), and the non-singlet singlet-like sector $P_{ns,+}$ is diagonal over
$\qty{T_3,T_8,T_{15},T_{24},T_{35}}$, and the non-singlet valence-like sector
$P_{ns,-}$ is diagonal over $\qty{V_3,V_8,V_{15},V_{24},V_{35}}$.
The respective distributions are given by their usual definition.

This Evolution Basis is isomorphic to our canonical choice
\begin{equation}
    \Fd \sim \Fd_{ev} = \vspan(g, \Sigma, V, T_{3}, T_{8}, T_{15}, T_{24}, T_{35}, V_{3}, V_{8}, V_{15}, V_{24}, V_{35})
\end{equation}
but, it is not a normalized basis. When dealing with intrinsic evolution, i.e.\
the evolution of \pdf{}s below their respective mass scale, the Evolution Basis
is not sufficient. In fact, for example, $T_{15} = u^{+} + d^{+} +
s^{+} - 3c^{+}$ below the charm threshold $\mu_c^2$ contains both running and static
distributions which need to be further disentangled.

We are thus considering a set of \enquote{Intrinsic Evolution Bases} $\Fd_{iev,
n_f}$, where we retain the intrinsic flavor distributions as basis vectors.
The basis definition depends on the number of light flavors $n_f$ and, e.g.\
for $n_f=4$, we find
\begin{equation}
    \Fd \sim \Fd_{iev,4} = \vspan(g, \Sigma_{(4)}, V_{(4)}, T_{3}, T_{8}, T_{15}, V_{3}, V_{8}, V_{15}, b^+, b^-, t^+, t^-)
\end{equation}
with $\Sigma_{(4)} = \sum\limits_{j=1}^4 q_j^+$ and $V_{(4)} =
\sum\limits_{j=1}^4 q_j^-$.

\subsection{Solution Strategies}
\label{sec:theory:solutions}
The formal solution of \cref{eq:dglap2} in terms of evolution kernel operators
$\vb {\tilde E}$ is given by
\begin{equation}
    \vb {\tilde E}(a_s \leftarrow a_s^0)  = \Pd \exp\qty[-\int\limits_{a_s^0}^{a_s} \frac{\bm{\gamma}(a_s')}{\beta(a_s')} \dd{a_s'} ]
    \label{eq:eko}
\end{equation}
with $\Pd$ the path-ordering operator. If the anomalous dimension $\bm{\gamma}$ is
diagonal in flavor space, i.e.\ it is in the non-singlet sector, it is always
possible to find an analytical solution to \cref{eq:eko}.  In the singlet
sector sector, however, this is only true at LO and to obtain a solution
beyond, we need to apply different approximations and solution strategies, on
which \eko{} offers currently eight implementations. For an actual comparison
of selected strategies, see \cref{sec:pheno:sols}.

\subsection{Matching at Thresholds}
\label{sec:theory:matching}
\eko{} can perform calculation in a fixed flavor number scheme (\ffns{}) where
the number of active or light flavors $n_f$ is constant. This means that both
the beta function $\beta^{(n_f)}(a_s)$ and the anomalous dimension
$\bm{\gamma}^{(n_f)}(a_s)$ in \cref{eq:dglap2} are constant with respect to
$n_f$.
However, this approximation is likely to fail either in the high energy region
$\muF^2 \to \infty$ for a small number of active flavors, or to fail in the low
energy region $\muF^2 \to \Lambda_{\text{QCD}}^2$ for a large number of active
flavors.

This can be overcome by using a variable flavor number scheme (\vfns{}) that
changes the number of active flavors when the scale $\muF^2$ crosses a
threshold $\mu_h^2$.
This then requires a matching procedure when changing the number of active
flavors, and for the \pdf{}s we find
\begin{equation}
    \tilde{\mathbf{f}}^{(n_f+1)}(\mu_{F,1}^2)= \tilde{\mathbf{E}}^{(n_f+1)}(\mu_{F,1}^2\leftarrow \mu_{h}^2) {\mathbf{R}^{(n_f)}} \tilde{\mathbf{A}}^{(n_f)}(\mu_{h}^2) \tilde{\mathbf{E}}^{(n_f)}(\mu_{h}^2\leftarrow \mu_{F,0}^2) \tilde{\mathbf{f}}^{(n_f)}(\mu_{F,0}^2)
    \label{eq:matching}
\end{equation}
where the superscript refers to the number of active flavors and we split the matching into two
parts: the perturbative operator matrix elements (\ome{}) $\tilde{\mathbf{A}}^{(n_f)}(\mu_{h}^2)$,
currently implemented at \nnlo{}~\cite{Buza_1998}, and an algebraic rotation ${\mathbf{R}^{(n_f)}}$ acting
only in the flavor space $\Fd$.

For backward evolution this matching has to be applied in the reversed order.
The inversion of the basis rotation matrices $\mathbf{R}^{(n_f)}$ is simple,
whereas this is not true for the \ome{} $\mathbf{\tilde A}^{(n_f)}$ especially
in case of \nnlo{} or higher order evolution.
In \eko{} we have implemented two different strategies to perform the inverse
matching: the first one is a numerical inversion, where the OMEs are inverted
exactly in Mellin space, while in the second method, called \texttt{expanded},
the matching matrices are inverted through a perturbative expansion in $a_s$,
given by:
\begin{align}
    \qty(\mathbf{\tilde A}^{(n_f)})_{exp}^{-1}(\mu_{h}^2) &= \mathbf{I} - a_s(\mu_{h}^2) \mathbf{\tilde A}^{(n_f),(1)} + a_s^2(\mu_{h}^2) \qty[ \mathbf{\tilde A}^{(n_f),(2)} - \qty(\mathbf{\tilde A}^{(n_f),(1)})^2 ] + O(a_s^3)
    \label{eq:invmatchingexp}
\end{align}
with $\mathbf{I}$ the identity matrix in flavor space.

\subsection{Running Quark Masses}
\label{sec:theory:msbarmass}
In the context of \pdf{} evolution, the most used treatment of heavy quarks masses are the pole masses,
where the physical values are specified as input and do not depend on any scale.
However for specific applications, such as the determination of \mhou{} due to heavy quarks contribution 
inside the proton~\cite{Ball:2016neh}, \msbar{} masses can also be used.
In particular, in~\cite{Alekhin:2010sv} it is found that higher order corrections on heavy quark production
in \dis{} are more stable upon scale variation when using the \msbar{} scheme.
\eko{} allows for this option as it is discussed briefly in the following paragraphs.

Whenever the initial condition for the mass is not given at a scale coinciding with
the mass itself (i.e. $\mu_{h,0} \neq m_{h,0}$, being $m_{h,0}$ the given initial condition
at the scale $\mu_{h,0}$),
\eko{} computes the scale at which the running mass $m_{h}(\mu_h^2)$ intersects
the identity function.
Thus for each heavy quark $h$ we solve:
\begin{equation}
    m_{\overline{MS},h}(m_h^2) = m_h
\end{equation}
The value $m_h(m_h)$ is then used as a reference to define the evolution thresholds.

The evolution of the \msbar{} mass is given by:
\begin{equation}
    m_{\overline{MS},h}(\mu_h^2) = m_{h,0} \exp\qty[ - \int\limits_{a_s(\mu_{h,0}^2)}^{a_s(\mu_h^2)} \frac{\gamma_m(a_s')}{\beta(a_s')} \dd{a_s'} ]
    \label{eq:msbarsolution}
\end{equation}
with $\gamma_m(a_s)$ the \qcd{} anomalous mass dimension available up to 
\nnnlo{}~\cite{Vermaseren:1997fq,Schroder:2005hy,Chetyrkin:2005ia}.

Note that to solve \cref{eq:msbarsolution} $a_s(\mu_R^2)$ must be evaluated in 
a \ffns{} until the threshold scales are known. Thus it is important
to start computing the \msbar{} masses of the quarks which are closer to the
the scale $\mu_{R,0}$ at which the initial reference value $a_s(\mu_{R,0}^2)$ is given. 

Furthermore, to find consistent solutions the boundary condition of the
\msbar{} masses must satisfy $m_h(\mu_h) \ge \mu_h$ for heavy quarks involving
a number of active flavors greater than the number of quark flavors $n_{f,0}$ at $\mu_{R,0}$, implying that we find
the intercept between the \rge{} and the identity in the forward direction ($m_{\overline{MS},h} \ge \mu_h$).
The opposite holds for scales related to fewer active flavors.

\section{Benchmarking and Validation}
\label{sec:pheno}
Although \eko{} is totally \pdf{} independent, for the sake of plotting
we choose NNPDF4.0~\cite{NNPDF:2021njg} as a default choice for
our plots, but for \cref{sec:pheno:bench} where we choose the toy \pdf{} of the
Les Houches Benchmarks~\cite{Giele:2002hx,Dittmar:2005ed}.
We show the gluon distribution $g(x)$ as a
representative member of the singlet sector and the valence distribution $V(x)$
as a representative member of the non-singlet sector.
Note that \pdf{}s in the same sector have mostly the same behavior, apart from
some specific regions (e.g.\ the $T_{15}$ distribution right after charm
matching).

\subsection{Benchmarks}
\label{sec:pheno:bench}
In this section we present the outcome of the benchmarks between \eko{} and similar 
available tools assuming different theoretical settings.

\subsubsection{Les Houches Benchmarks}
\eko{} has been compared with the benchmark tables
given in \cite{Giele:2002hx,Dittmar:2005ed}.
We find a good match except for a list of typos which we list here:
\begin{itemize}
    \item in table head in \cite{Giele:2002hx} should be $2xL_+ = 2x(\bar u + \bar d)$
    \item in the head of table 1: the value for $\alpha_s$ in \ffns{} is wrong (as pointed out and corrected in \cite{Dittmar:2005ed})
    \item in table 3, part 3 of \cite{Giele:2002hx}: $xL_-(x=10^{-4}, \muF^2 = \SI{1e4}{\GeV^2})=1.0121\cdot 10^{-4}$ (wrong exponent) and
          $xL_-(x=0.1, \mu_F^2 = \SI{1e4}{\GeV^2})=9.8435\cdot 10^{-3}$ (wrong exponent)
    \item in table 15, part 1 of \cite{Dittmar:2005ed}: $xd_v(x=10^{-4}, \mu_F^2 = \SI{1e4}{\GeV^2}) = 1.0699\cdot 10^{-4}$ (wrong exponent) and
          $xg(x=10^{-4}, \mu_F^2 = \SI{1e4}{\GeV^2}) = 9.9694\cdot 10^{2}$ (wrong exponent)
\end{itemize}
Some of these typos have been already reported in \cite{Diehl:2021gvs}.

\begin{figure}
    \centering
    \includegraphics[width=\textwidth,height=.24\textheight]{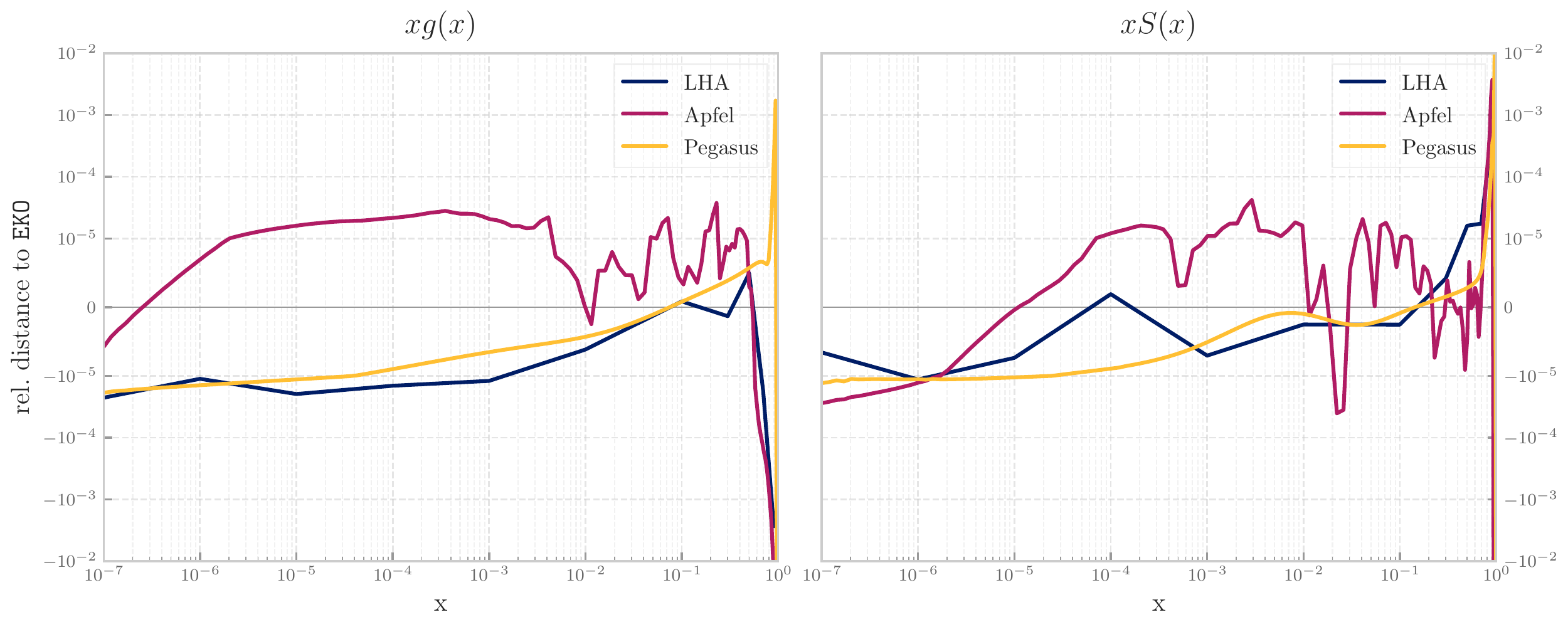}
    \includegraphics[width=\textwidth,height=.24\textheight]{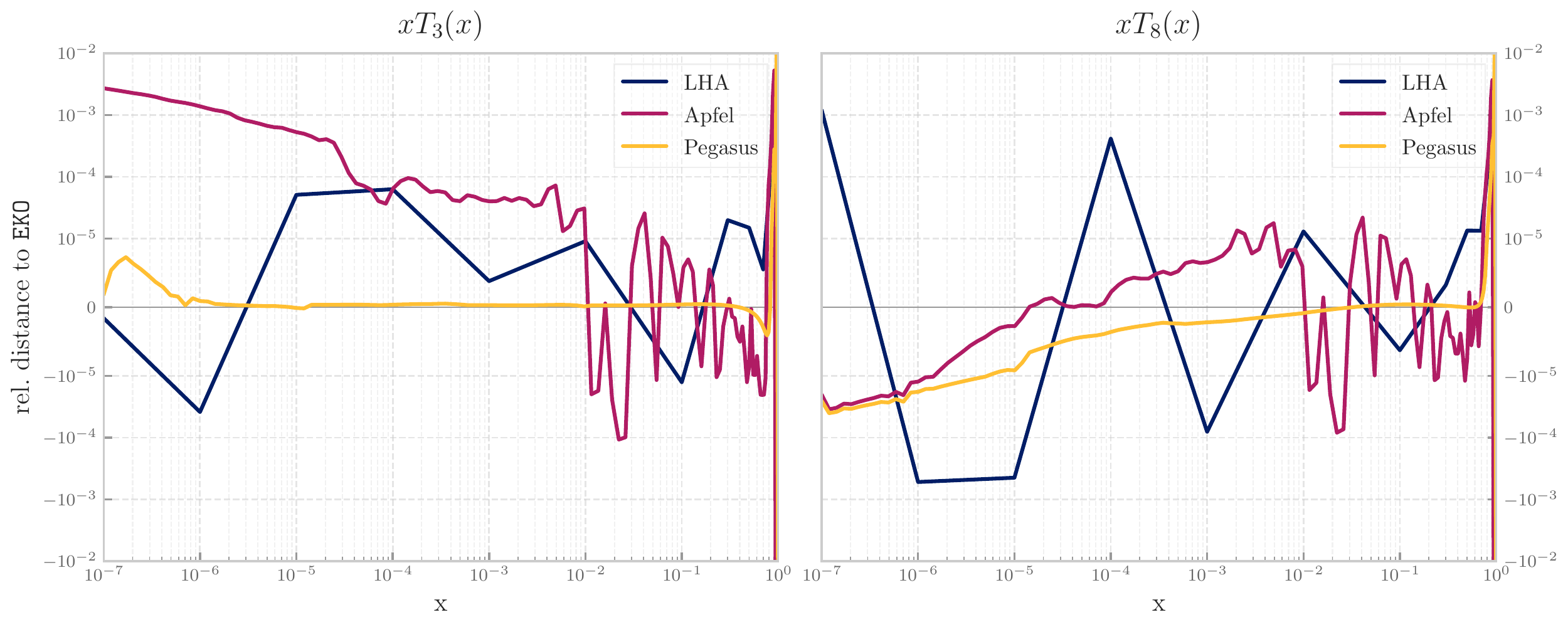}
    \includegraphics[width=\textwidth,height=.24\textheight]{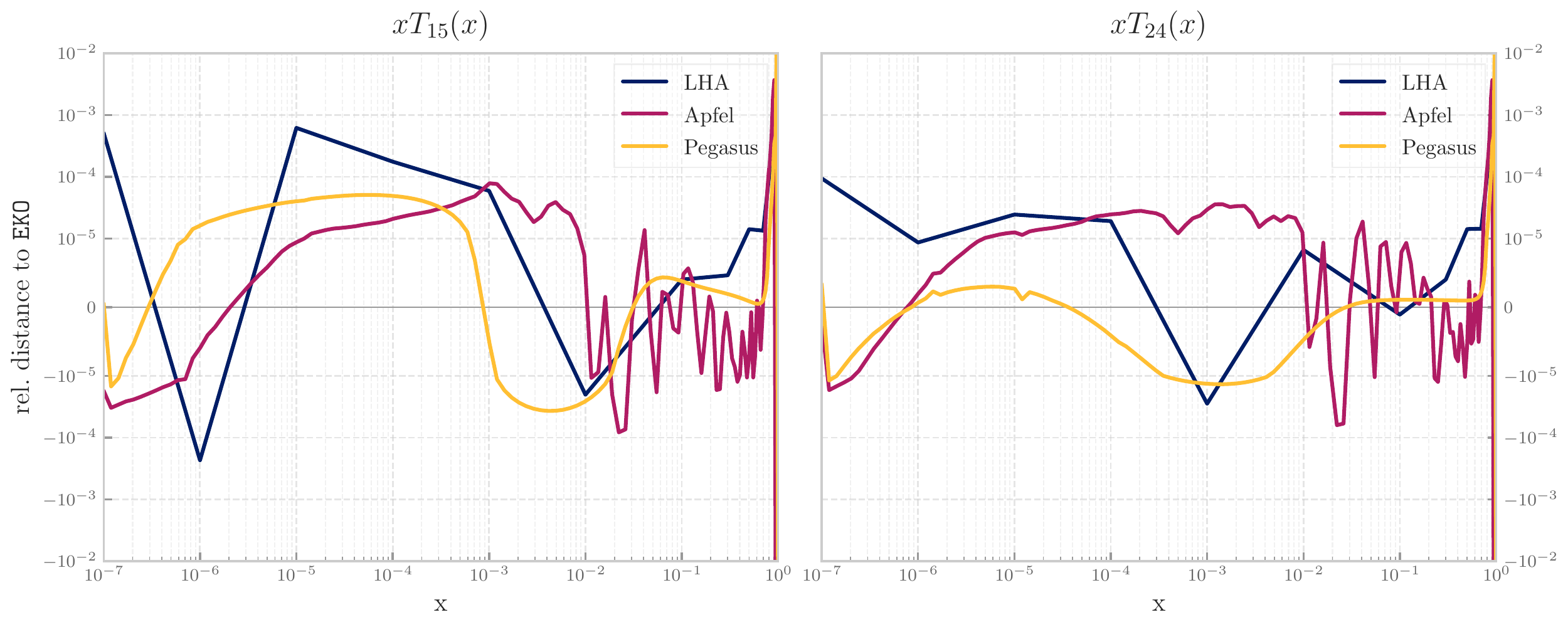}
    \includegraphics[width=\textwidth,height=.24\textheight]{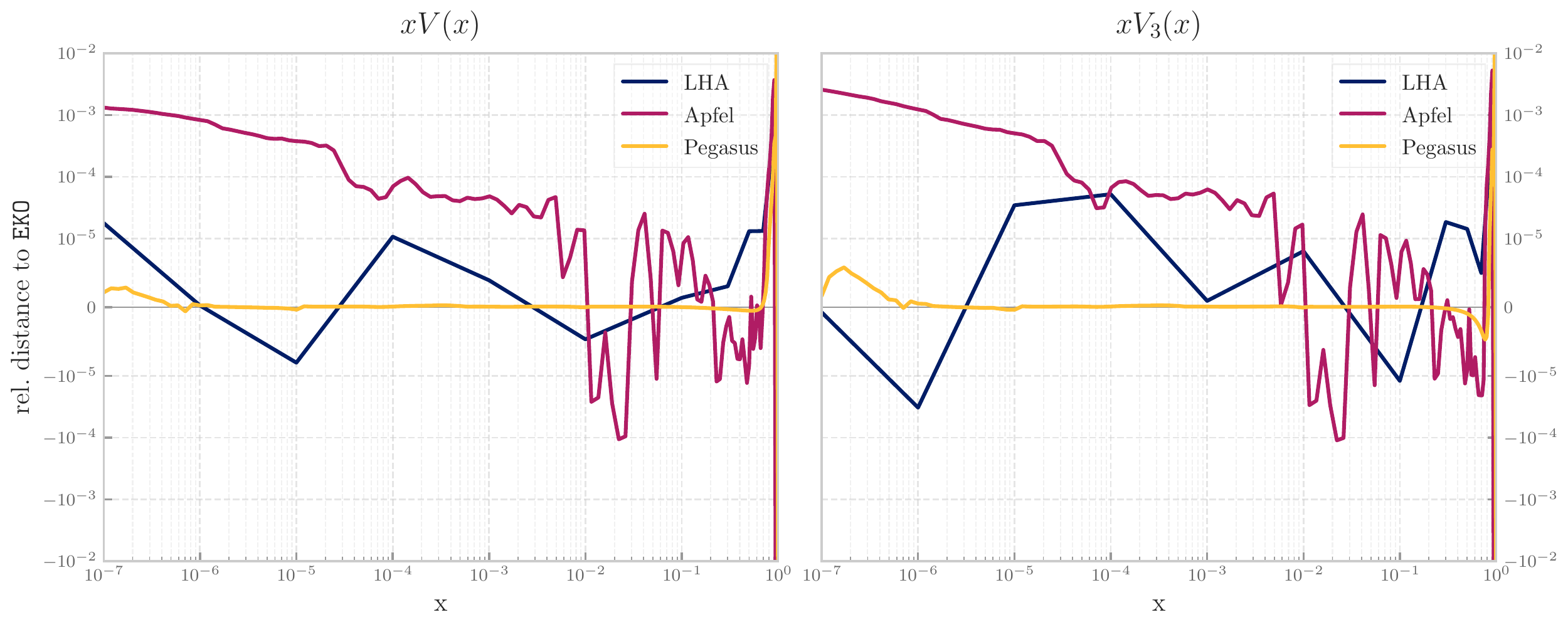}
    \caption{Relative differences between 
        the outcome of \nnlo{} \qcd{} evolution
        as implemented in \eko{} and the
        corresponding results from \cite{Dittmar:2005ed}, \apfel{}~\cite{Bertone:2013vaa} and \pegasus{}~\cite{Vogt:2004ns}.
        We adopt the settings of the Les Houches \pdf{} evolution benchmarks~\cite{Giele:2002hx,Dittmar:2005ed}.}
    \label{fig:LHAbench}
\end{figure}

In \cref{fig:LHAbench} we present the results of the \vfns{} benchmark at \nnlo{}, where a
toy \pdf{} is evolved from $\mu_{F,0}^2=\SI{2}{\GeV^2}$ up to $\mu_{F}^2=\SI{1e4}{\GeV^2}$
with equal values of the factorization and renormalization scales $\muF=\muR$.
For completeness, we display the singlet $S(x)$ and gluon $g(x)$ distribution (top), the singlet-like $T_{3,8,15,24}(x)$ (middle)
and the valence $V(x)$, valence-like $V_{3}(x)$ (bottom) along with
the results from \apfel{} and \pegasus{}. We find an overall agreement at the level of $O(10^{-3})$.

\subsubsection{APFEL}
\apfel{} \cite{Bertone:2013vaa} is one of the most extensive tool aimed to
\pdf{}  evolution and \dis{} observables' calculation. It is provided as a
Fortran library, and it has been used by the NNPDF collaboration up to
NNPDF4.0~\cite{NNPDF:2021njg}.

\apfel{} solves \dglap{} numerically in $x$-space, sampling the evolution
kernels on a grid of points up to \nnlo{} in \qcd{}, with \qed{} evolution also
available at \lo{}.
By construction this method is \pdf{} dependent and the code is automatically
interfaced with \lhapdf{}~\cite{Buckley:2014ana}. For specific application,
the code offers also the possibility to retrieve the evolution operators
with a dedicated function.

The program supplies three different solution strategies, with various theory
setups, including scale variations and \msbar{} masses.

\begin{figure}
    \includegraphics[width=\linewidth]{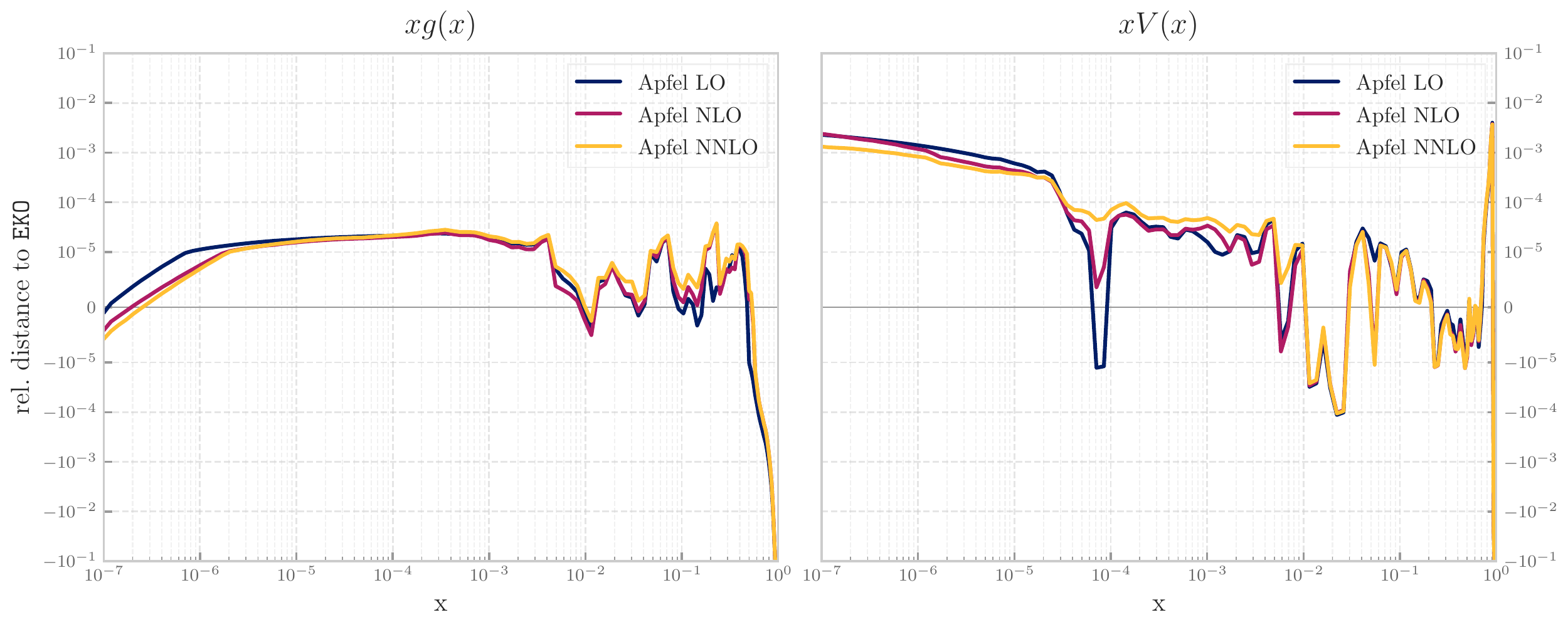}
    \caption{Relative differences between the outcome of evolution as
        implemented in \eko{} and the corresponding results from \apfel{} at
        different perturbative orders.  We adopt the same settings of
        \cref{fig:LHAbench}.}
    \label{fig:Apfelbench_pto}
\end{figure}

The stability of our benchmark at different perturbative orders is presented in \cref{fig:Apfelbench_pto},
using the settings of the Les Houches \pdf{} evolution benchmarks~\cite{Giele:2002hx,Dittmar:2005ed}.
The accuracy of our comparison is not affected by the increasing complexity
of the calculation.

\subsubsection{PEGASUS}
\pegasus{}~\cite{Vogt:2004ns} is a Fortran program aimed for \pdf{} evolution.
The program solves \dglap{} numerically in $N$-space up to \nnlo{}.
\pegasus{} can only deal with \pdf{}s given as a fixed functional form and is
not interfaced with \lhapdf{}.

As shown in \cref{fig:LHAbench}, the agreement of \eko{} with this tool is better than with \apfel{},
especially for valence-like quantities, for which an exact solution is possible, where we reach
$\mathcal{O}(10^{-6})$ relative accuracy.
This is expected and can be traced back to the same \dglap{} solution strategy in Mellin space.

\begin{figure}
    \includegraphics[width=\linewidth]{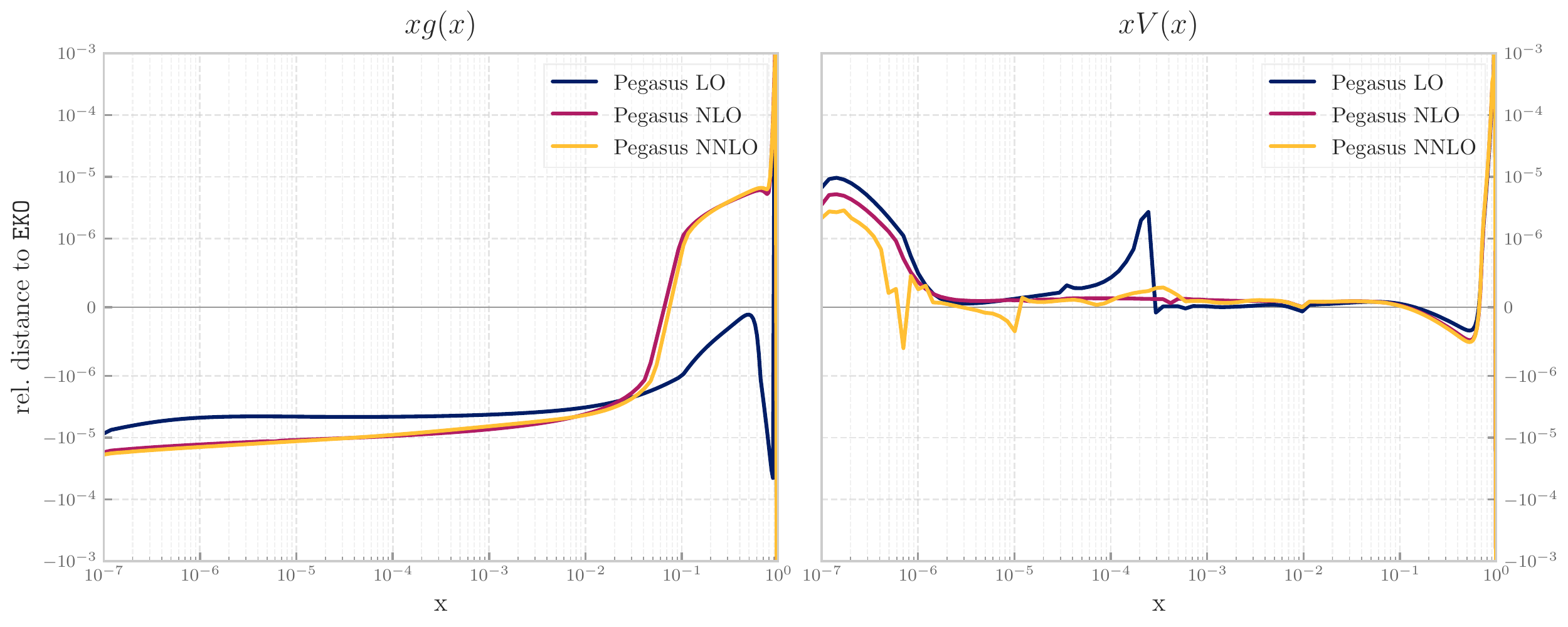}
    \caption{Same of \cref{fig:Apfelbench_pto}, now comparing to \pegasus{}~\cite{Vogt:2004ns}.
        \label{fig:Pegasusbench_pto} }
\end{figure}

Similarly to the \apfel{} benchmark, we assert that the precision of our benchmark with \pegasus{} is not affected
by the different \qcd{} perturbative orders, as visible in \cref{fig:Pegasusbench_pto}.
As both, \apfel{} and \pegasus{}, have been benchmarked against
\hoppet{}~\cite{Salam:2008qg} and \qcdnum{}~\cite{Botje:2010ay} we conclude
to agree also with these programs.

\subsection{Solution Strategies}
\label{sec:pheno:sols}
As already mentioned in \cref{sec:theory:solutions}, due to
the coupled integro-differential structure of \cref{eq:dglap}, 
solving the equations requires in practice some approximations to which we refer as different
solution strategies. \eko{} currently implements 8 different strategies,
corresponding to different approximations.
Note that they may differ only by the strategy in a specific sector,
such as the singlet or non-singlet sector. All provided strategies
agree at fixed order, but differ by higher order terms.

\begin{figure}
    \centering
    \includegraphics[width=\textwidth]{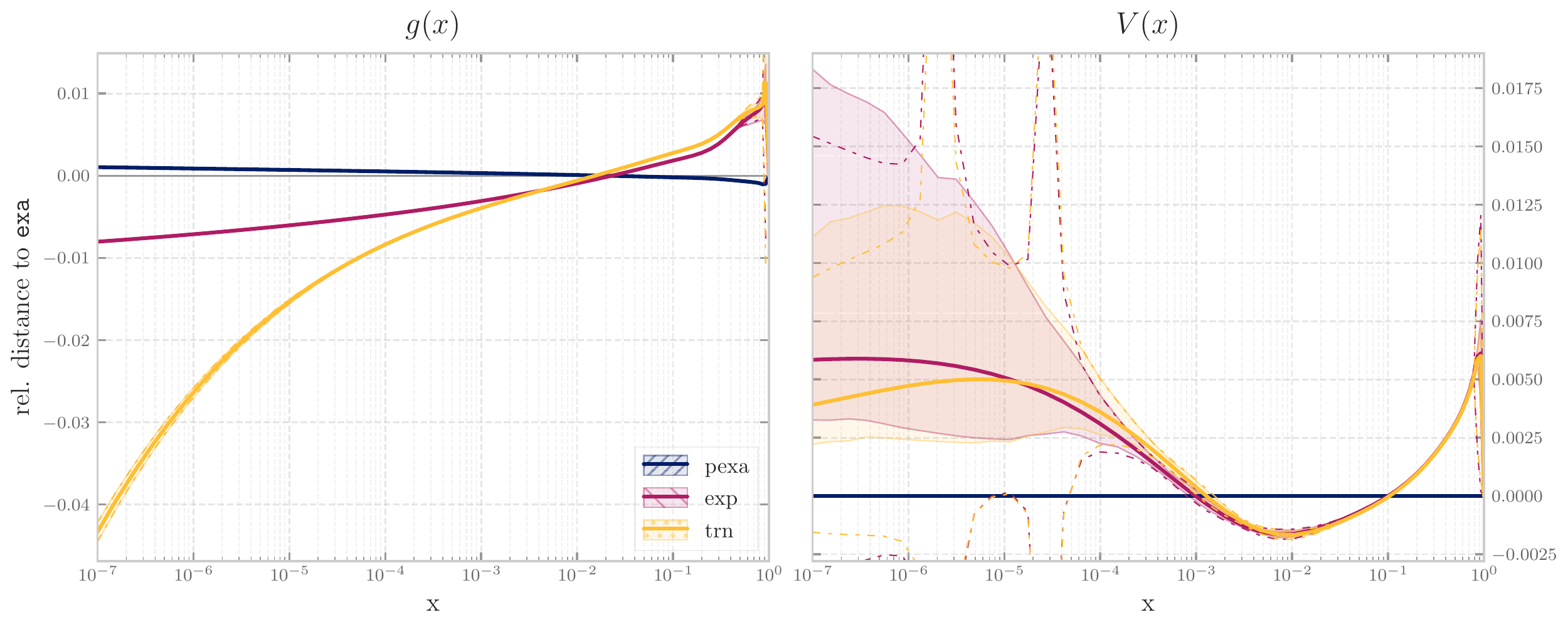}
    \caption{Compare selected solutions strategies, with respect to the
        \texttt{iterated-exact} (called \texttt{exa} in label) one. In
        particular: \texttt{perturbative-exact} (\texttt{pexa}) (matching
        the reference in the non-singlet sector),
        \texttt{iterated-expanded} (\texttt{exp}), and \texttt{truncated}
        (\texttt{trn}).
        The distributions are evolved in $\muF^2=\SI[parse-numbers=false]{1.65^2\to 10^4}{\GeV^2}$.}
    \label{fig:solutions}
\end{figure}

In \cref{fig:solutions} we show a comparison of a selected list of
solution strategies\footnote{For the full list of available solutions and a
detailed descriptions see the
\href{https://eko.readthedocs.io/en/latest}{online documentation}.}:

\begin{itemize}
    \item \texttt{iterate-exact}:
        In the non-singlet sector we take the analytical solution
        of \cref{eq:dglap2} up to the order specified.
        In the singlet sector we split the evolution path into segments
        and linearize the exponential in each segment~\cite{Bonvini:2012sh}.
        This provides effectively a straight numerical solution of \cref{eq:dglap2}.
        In \cref{fig:solutions} we adopt this strategy as a reference.
    \item \texttt{perturbative-exact}:
        In the non-singlet sector it coincides with \texttt{iterate-exact}.
        In the singlet sector we make an ansatz to determine the solution as a
        transformation $\vb{U}(a_s)$ of the \lo{} solution~\cite{Vogt:2004ns}. We then
        iteratively determine the perturbative coefficients of $\textbf{U}$.
    \item \texttt{iterate-expanded}:
        In the singlet sector we follow the strategy of \texttt{iterate-exact}.
        In the non-singlet sector we expand \cref{eq:dglap2} first to the order
        specified, before solving the equations.
    \item \texttt{truncated}: 
        In both sectors, singlet and non-singlet, we make an ansatz to determine the solution as a
        transformation $\vb{U}(a_s)$ of the \lo{} solution and
        then expand the transformation $\vb U$ up to the order specified.
        Note that for programs using $x$-space this strategy is difficult
        to pursue as the \lo{} solution is kept exact and only the transformation
        $\vb U$ is expanded.
\end{itemize}

The strategies differ most in the small-$x$ region where the \pdf{} evolution is
enhanced and the treatment of sub-leading corrections become relevant.
This feature is seen most prominently in the singlet sector between
\texttt{iterate-exact} (the reference strategy) and \texttt{truncated}.
In the non-singlet sector the distributions also vanish for small-$x$
and so the difference gets artificially enhanced.
This is eventually the source of the spread for the valence distribution $V(x)$
making it more sensitive to the initial \pdf{}.

\paragraph{PDF plots} The \pdf{} plot shown in \cref{fig:solutions} contains
multiple elements, and its layout is in common with
\cref{fig:interpolation,fig:pdfmatching}.

All the different entries corresponds to different theory settings, and they
are normalized with respect to a reference theory setup
(e.g.\ in \cref{fig:solutions} the \texttt{iterative-exact} strategy)
and the lines correspond to the relative difference.

Furthermore, an envelope and dashed lines are displayed.
To obtain them, the full \pdf{} set is evolved, replica by replica, for each
configuration (corresponding to a single evolution operator, that is applied to
each replica in turn).
Then ratios are taken between corresponding evolved replicas, to highlight
the \pdf{} independence of \eko{} rather then any specific set-related features.
The upper and lower borders of the envelope correspond respectively
to the $0.16$ and $0.84$ quantiles of the replicas set,
while the dashed lines correspond to one standard deviation.

\subsection{Interpolation}
\label{sec:pheno:interp}
To bridge between the desired $x$-space input/output and the internal
Mellin representation, we do a Lagrange-Interpolation as sketched in
\cref{sec:theory:interpolation}
(and detailed in the \href{https://eko.readthedocs.io/en/latest/}{online documentation}).
We recommend a grid of at least 50 points with
linear scaling in the large-$x$ region ($x \gtrapprox 0.1$) and with logarithmic
scaling in the small-$x$ region and an interpolation of degree four.
Also the grids determined by \amcfast~\cite{Bertone:2014zva} perform
sufficiently well for specific processes.

\begin{figure}
    \begin{center}
    \includegraphics[width=\textwidth]{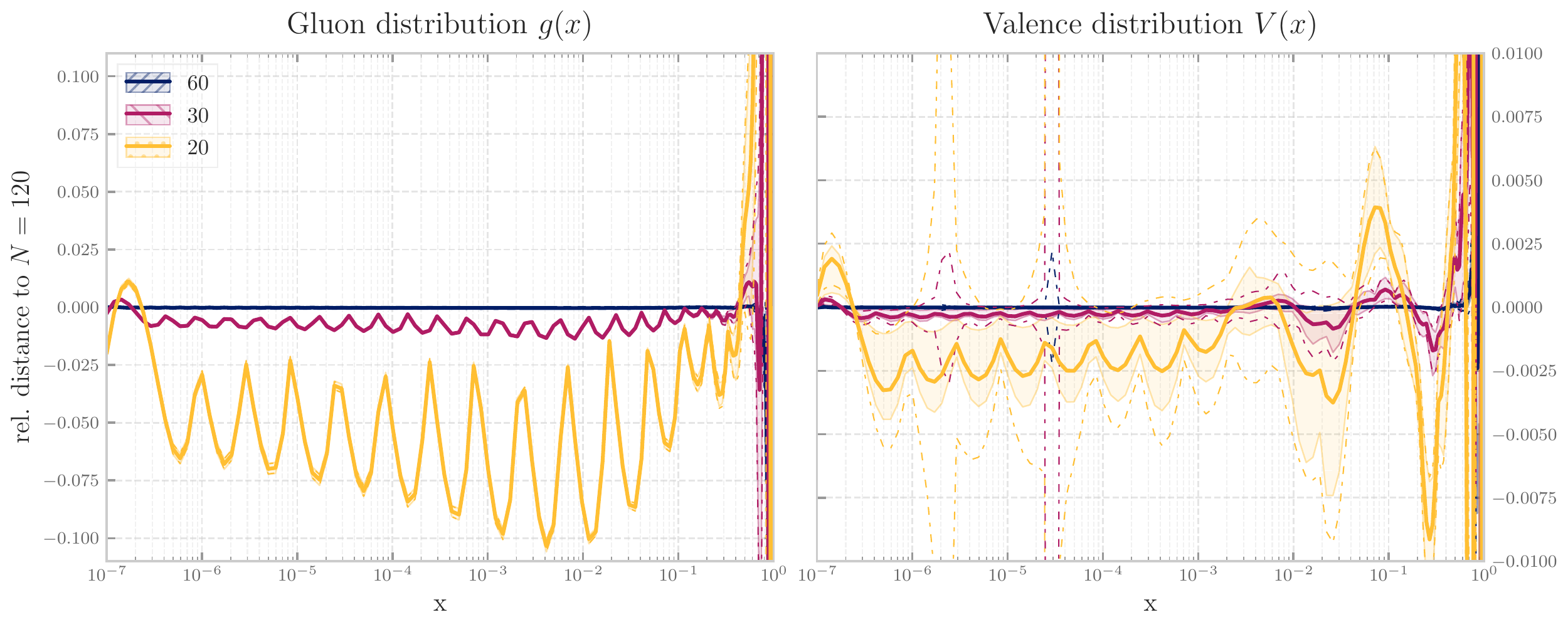}
    \end{center}
    \caption{Relative differences between 
        the outcome of \nnlo{} \qcd{} evolution
        as implemented in \eko{} with 20, 30, and 60 points to 120
        interpolation points respectively.
        \label{fig:interpolation} }
\end{figure}

For a first qualitative study we show in \cref{fig:interpolation} a
comparison between an increasing number of interpolation points
distributed according to \cite[Eq. 2.12]{Carrazza_2020}.
The separate configurations are converging to the solution with the
largest number of points. Using 60 interpolation points is almost
indistinguishable from using 120 points (the reference configuration in the plot).
In the singlet sector (gluon) the convergence is
significantly slower due to the more involved solution strategies and,
specifically, the oscillating behavior is caused due to these difficulties.
The spikes for $x\to 1$ are not relevant since the \pdf{}s are intrinsically
small in this region ($\vb f\to 0$) and thus small numerical differences
are enhanced.

Also note that the results of \cref{sec:pheno:bench} (i.e.\ \cref{fig:LHAbench,fig:Apfelbench_pto,fig:Pegasusbench_pto}) confirm that
the interpolation error can be kept below the benchmark accuracy.

\subsection{Matching}
\label{sec:pheno:match}
We refer to the specific value of the factorization scale at which the number
of active flavors is changing from $n_f$ to $n_f+1$ (or vice-versa) as the
threshold $\mu_h$. Although this value usually coincides with the respective
quark mass $m_h$, \eko{} implements the explicit expressions when the two
scales do not match. This variation can be used to estimate \mhou{}~\cite{AbdulKhalek:2019ihb}.

\begin{figure}
    \centering
    \includegraphics[width=\textwidth]{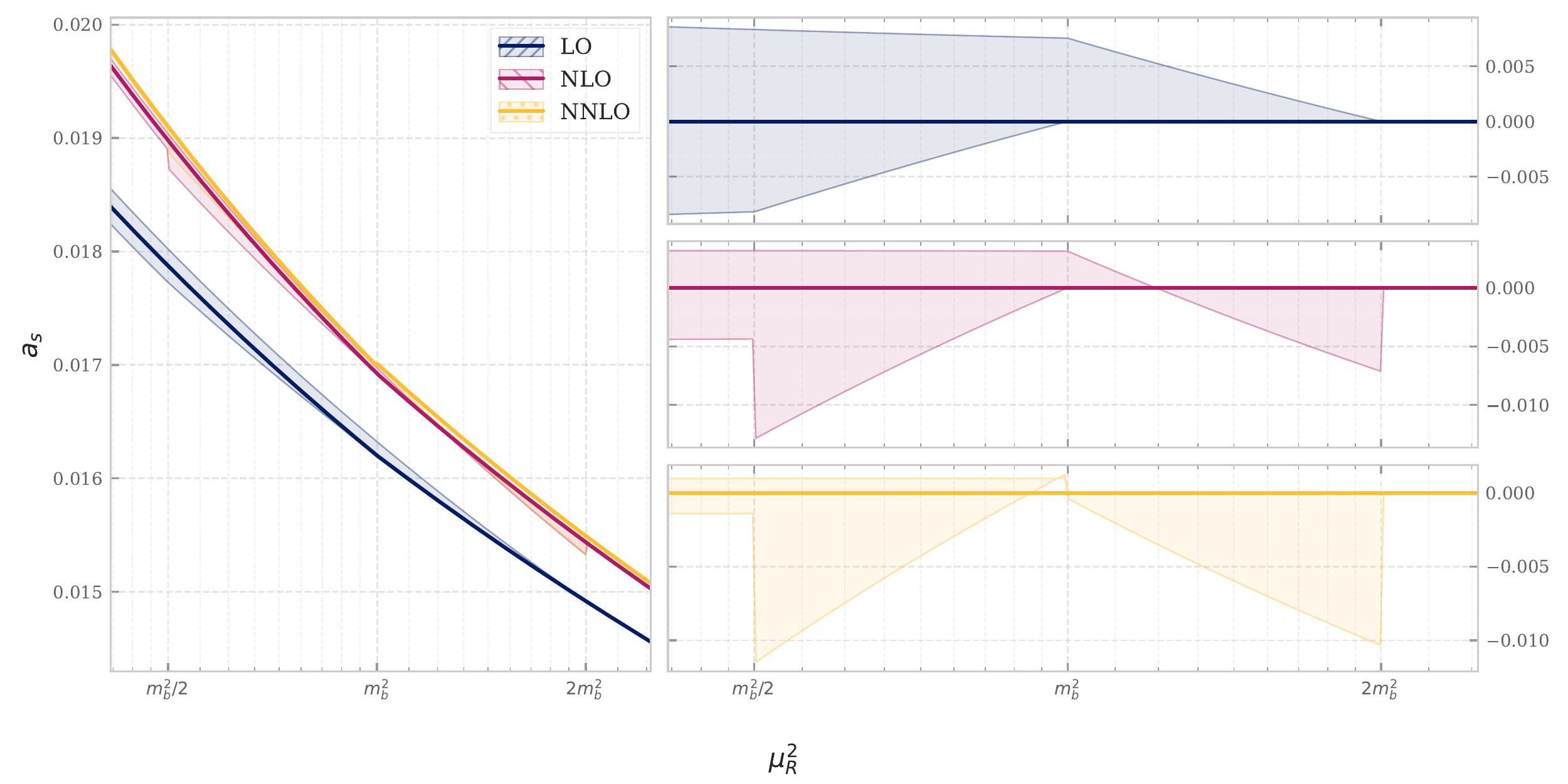}
    \caption{Strong coupling evolution $a_s(\mu_R^2)$ at \lo{},
    \nlo{} and \nnlo{}
    respectively with the bottom matching $\mu_b^2$ at $1/2, 1,$ and $2$ times the
    bottom mass $m_b^2$ indicated by the band. In the left panel we show the absolute
    value, while on the right we show the ratio towards the central scale choice.
     \label{fig:asmatching}}
\end{figure}

In \cref{fig:asmatching} we show the strong coupling evolution $a_s(\mu_R^2)$
around the bottom mass with the bottom threshold $\mu_b^2$ eventually not coinciding
with the respective bottom quark mass $m_b^2$.
The dependency on the \lo{} evolution is only due to the change of active
flavor in the beta function ($\beta_0 = \beta_0(n_f)$), which can be seen in
the ratio plot by the continuous connections of the lines.
At \nlo{} evolution the matching condition already becomes discontinuous for
$\mu_h^2 \neq m_h^2$, represented in the ratio plot by the offset for the
matched evolution. 
The matching for the \nnlo{} evolution~\cite{Chetyrkin:2005ia,Schroder:2005hy}
is intrinsically discontinuous, which is indicated in the ratio plot by the
discrete jump at the central scale $\mu_R^2 = m_b^2$.
For $\mu_R^2 > 2m_b^2$ the evolution is only determined by the reference value
$a_s(m_Z^2)$ and the perturbative evolution order.
For $\mu_R^2 < m_b^2/2$ we can observe the perturbative convergence as the
relative difference shrinks with increasing orders.
Since it is converging, the effect of the matching condition should cancel more
and more exactly with the difference in running, but the magnitude of both is
increasing with the order, since the perturbative expansion of the beta
function $\beta(a_s)$ is a same sign series.

\begin{figure}
    \centering
    \includegraphics[width=\textwidth]{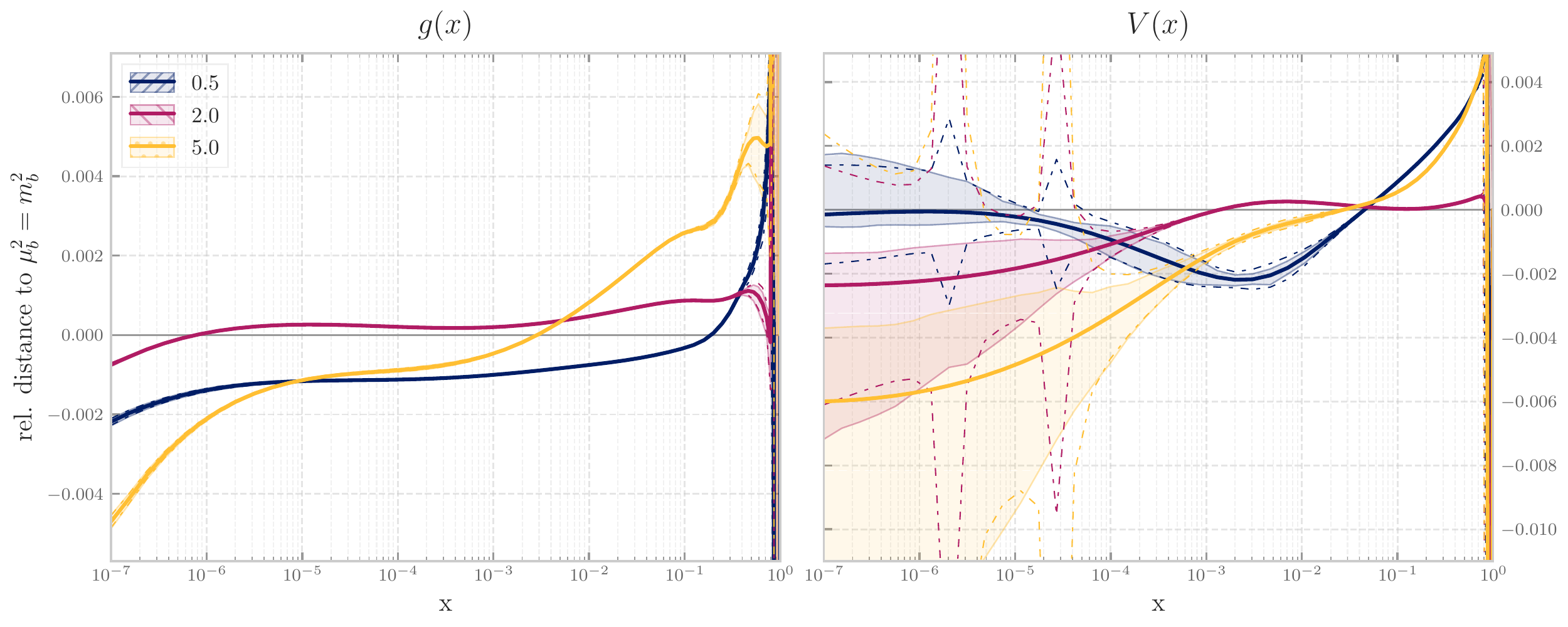}
    \caption{Difference of \pdf{} evolution with the bottom matching $\mu_b^2$ at $1/2, 2,$ and
        $5$ times the bottom mass $m_b^2$ relative to $\mu_b^2 = m_b^2$. Note
        the different scale for the two distributions.  All evolved in
        $\muF^2=\SI[parse-numbers=false]{1.65^2\to 10^4}{\GeV^2}$.}
    \label{fig:pdfmatching}
\end{figure}

In \cref{fig:pdfmatching} we show the relative difference for the \pdf{} evolution with
threshold values $\mu_h^2$ that do not coincide with the respective heavy
quark masses $m_h^2$. When matching at a lower scale the difference is
significantly more pronounced as the evolution includes
a region where the strong coupling varies faster. When dealing 
with $\mu_h^2 \neq m_h^2$ the \pdf{} matching conditions become discontinuous
already at \nlo{}~\cite{Buza_1998}. These contributions are also available in
\apfel{}~\cite{Bertone:2013vaa}, but not in \pegasus{}~\cite{Vogt:2004ns} and although they are present in the code
of \qcdnum{}~\cite{Botje:2010ay} they can not be accessed there.
For the study in \cite{Ball:2022qks} we also implemented the \pdf{} matching at
\nnnlo{}~\cite{Bierenbaum:2009zt,Bierenbaum:2009mv,Ablinger:2010ty,Ablinger:2014vwa,Ablinger:2014uka,Behring:2014eya,Ablinger_2014,Ablinger_2015,Blumlein:2017wxd}.

\subsection{Backward}
\label{sec:pheno:back}
\begin{figure}
    \begin{center}
    \includegraphics[width=\textwidth]{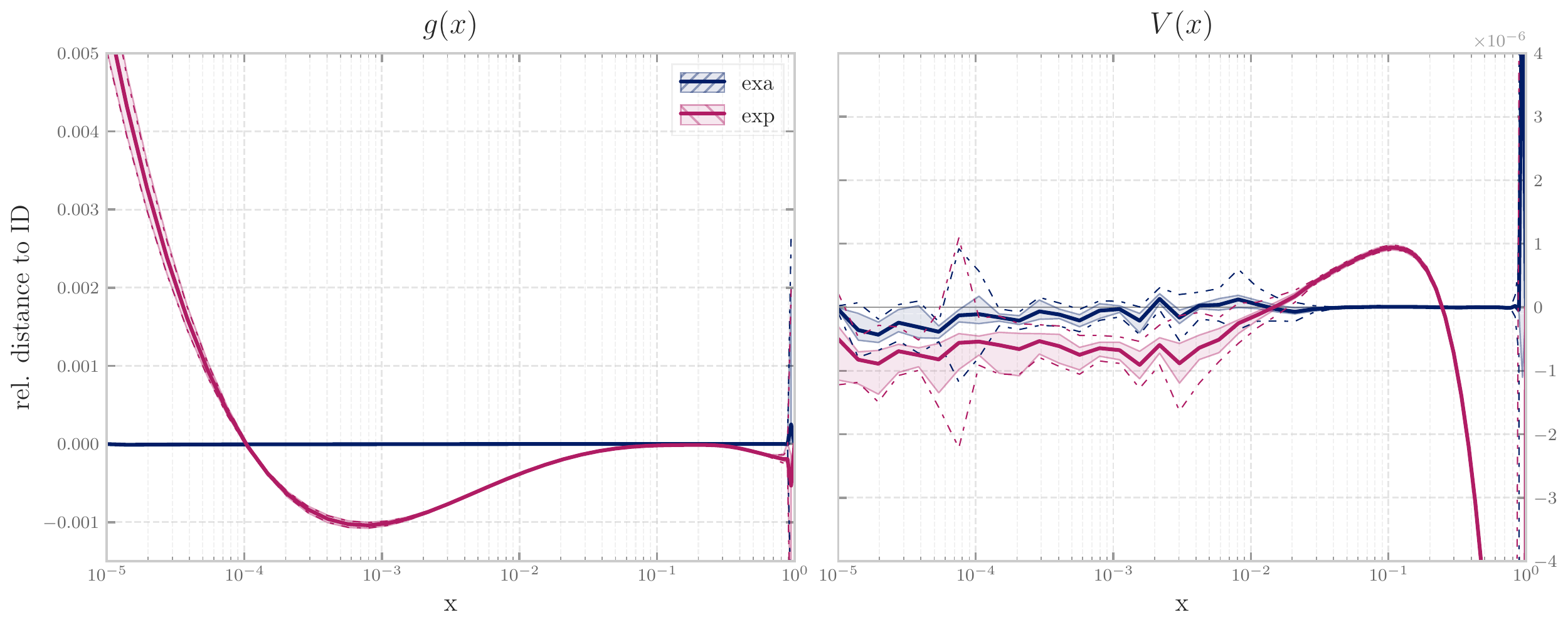}
    \end{center}
    \caption{Relative distance of the product of two opposite \nnlo{} \ekos{}
        and the identity matrix, in case of exact inverse and expanded
        matching (see \cref{eq:invmatchingexp}) when crossing the bottom
        threshold scale $\mu_{b}^2=\SI[parse-numbers=false]{4.92^2}{\GeV^2}$. In particular the lower scale is chosen $\muF^2=\SI[parse-numbers=false]{4.90^2}{\GeV^2}$, 
        while the upper is equal to $\muF^2=\SI[parse-numbers=false]{4.94^2}{\GeV^2}$, 
        \label{fig:closure_test}
    }
\end{figure}

As a consistency check we have performed a closure test verifying that after
applying two opposite \ekos{} to a custom initial condition
we are able to recover the initial \pdf{}. Specifically, the product of the
two kernels is an identity both in flavor and momentum space up to
the numerical precision. The results are shown in \cref{fig:closure_test} in case of \nnlo{} evolution
crossing the bottom threshold scale $\mu_{F}=m_{b}$. The differences between
the two inversion methods are more visible for singlet-like quantities,
because of the non-commutativity of the matching matrix $\tilde{\mathbf{A}}_{S}^{(n_f)}$.  

Special attention must be given to the heavy quark distributions which are
always treated as intrinsic, when performing backward evolution.
In fact, if the initial \pdf{} (above the mass threshold) contains an intrinsic contribution, this has to be evolved
below the threshold otherwise momentum sum rules can be violated.
This intrinsic component is then scale independent and fully decoupled
from the evolving (light) \pdf{}s.
On the other hand, if the initial \pdf{} is purely perturbative, it vanishes
naturally below the mass threshold scale after having applied the
inverse matching.
In this context, \eko{} has been used in a recent study to determine, for the first time,
the intrinsic charm content of the proton~\cite{Ball:2022qks}.

\subsection{\msbar{} masses}
\label{sec:pheno:msbarmass}
\begin{figure}
    \centering
    \includegraphics[width=0.47\linewidth]{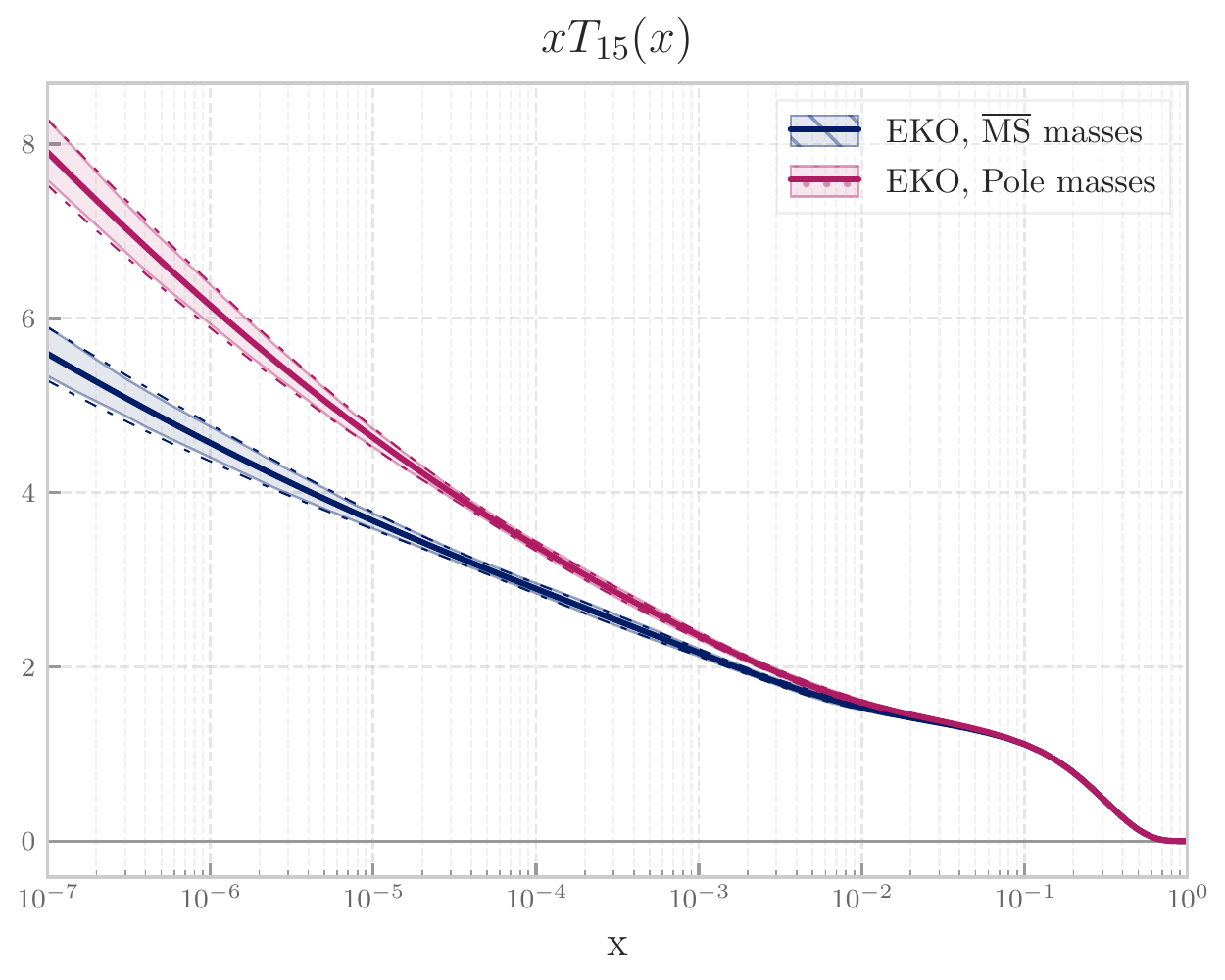}
    \includegraphics[width=0.47\linewidth]{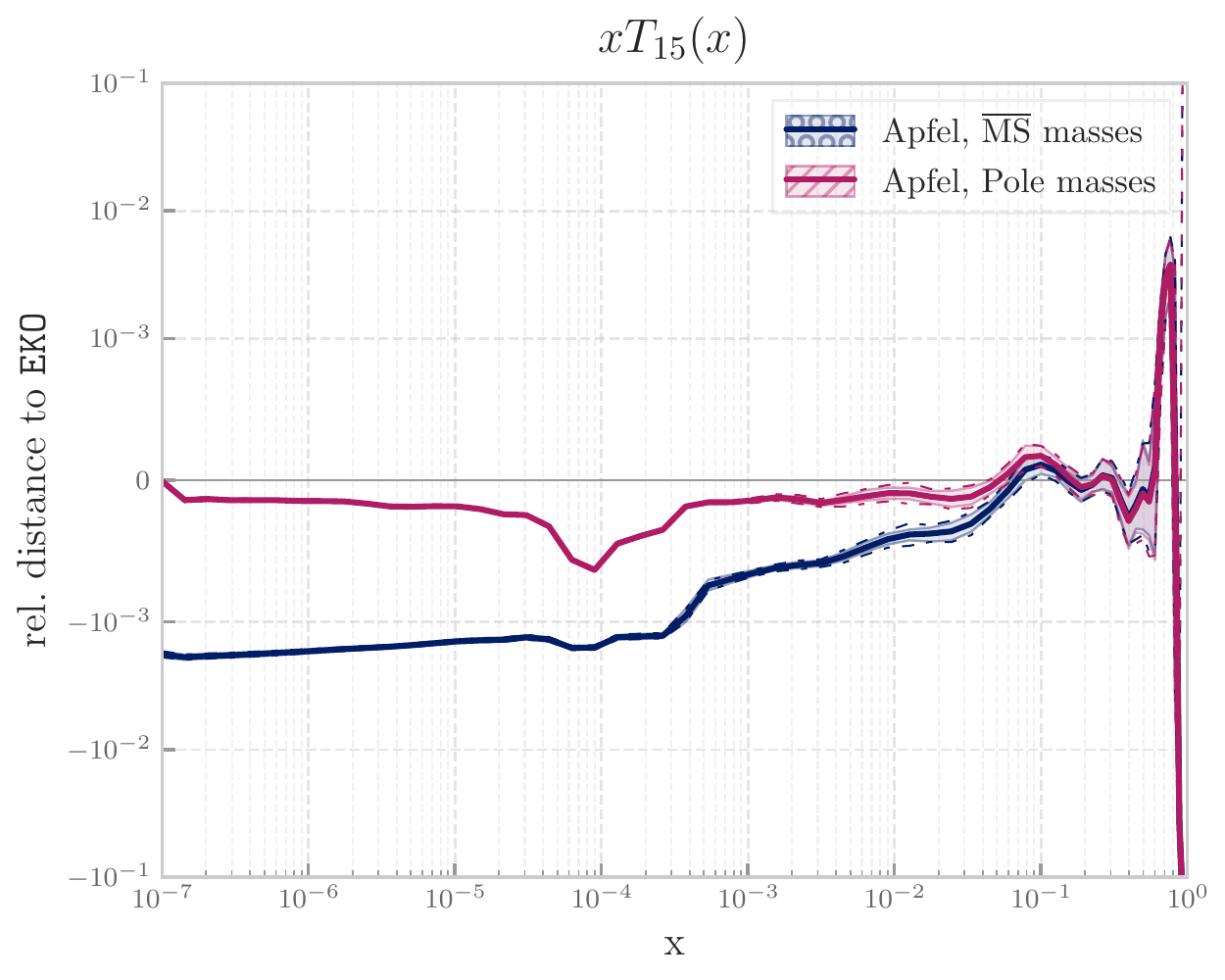}
    \caption{(left) The NNPDF4.0 perturbative charm distribution
        $T_{15}(x)$~\cite{NNPDF:2021njg} with \msbar{} and pole masses \nnlo{}
        evolution when running on $\muF^2=\SI[parse-numbers=false]{1\to
        10^4}{\GeV^2}$.  (right) Relative difference to \eko{} for the same run
        with \apfel{}~\cite{Bertone:2013vaa}.}
     \label{fig:MSbarbench}
\end{figure}

In \cref{fig:MSbarbench} we investigate the effect of adopting a running mass
scheme onto the respective \pdf{} sets. The left panel shows the $T_{15}(x)$
distribution obtained from the NNPDF4.0 perturbative charm
determination~\cite{NNPDF:2021njg} using the pole mass scheme and the \msbar{}
scheme, respectively.
The distributions have been evolved on $\muF^2=\SI[parse-numbers=false]{1\to
10^4}{\GeV^2}$.
The mass reference values are taken from
\cite{LHCHiggsCrossSectionWorkingGroup:2016ypw}, with the \msbar{}
boundary condition on the charm mass given as $m_c(\mu_m=\SI{3}{\GeV}) =
\SI{0.986}{\GeV}$, leading to $m_c(m_c) = \SI{1.265}{\GeV}$, while the charm
pole mass is $m^{\rm pole}_{c}\approxeq\SI{1.51}{\GeV}$~\cite{NNPDF:2021njg}.
The major differences are visible in the low-x region where the \dglap{}
evolution is faster and the differences between the charm mass treatment are
enhanced: an higher value of the charm mass increases the singlet like
distribution $T_{15}(x)$.
For the sake of comparison, in the right panel, we plot the relative distance
to our result when comparing with the \apfel{}~\cite{Bertone:2013vaa}
implementation.
As expected the pole mass results are closer due to the same value of the charm
mass, while the \msbar{} results have a slightly bigger discrepancy which
remains in all the $x$-range around $1\permil$ accuracy.

\begin{figure}
    \centering
    \includegraphics[width=\textwidth]{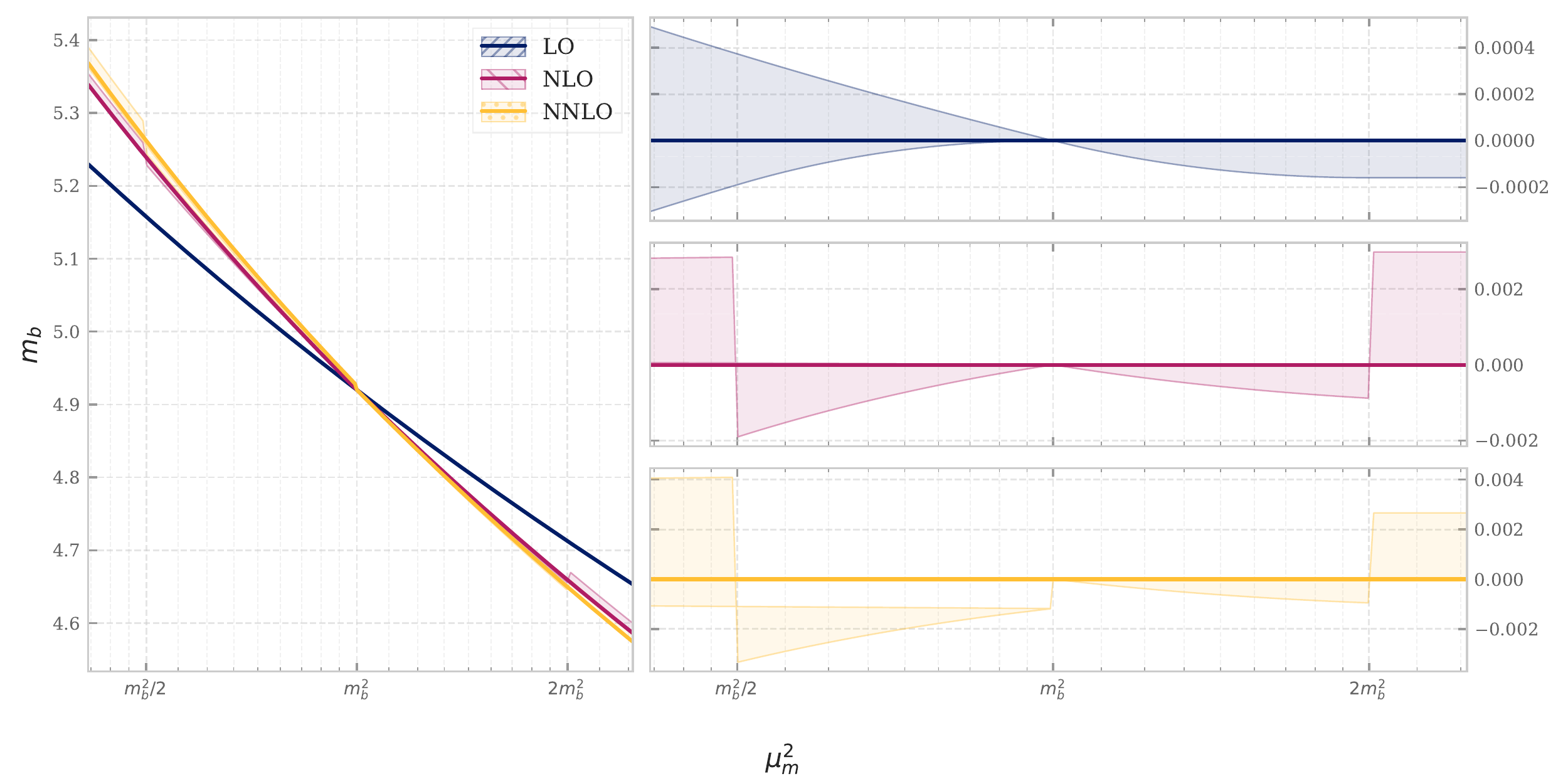}
    \caption{Running of the bottom quark mass $m_b(\mu_m^2)$ for different threshold
        ratios, similar to \cref{fig:asmatching}.
        The plot shows how the different choices of matching scales affect the
        running in the matching region (and slightly beyond) at \lo{}, \nlo{},
        and \nnlo{}.
        The border condition for the running has been chosen at $m_b(m_b) =
        \SI{4.92}{GeV}$, as it is clear from the plot, since it is the
        intersection point of all the curves shown.}
    \label{fig:runningmasses}
\end{figure}

In \cref{fig:runningmasses} we show the evolution of the \msbar{} bottom mass
$m_b(\mu_m^2)$ using different matching scales $\mu_b^2$ equal to $1/2,1$ and
$2$ times the mass $m_b^2$, for each perturbative order (\lo{}, \nlo{}, and
\nnlo{}).
The curve for $\mu_b^2 = m_b^2$ has been plotted as the central one (bold),
while the other two are used as the upper and lower borders of the shaded area
(according to their value, point by point).
The reference value $m_b(\mu_{b,0}^2)$, has been chosen equal for the
three curves, and it has been chosen at $m_b(m_b) = \SI{4.92}{GeV}$.
For this reason, above the central matching point $\mu_m^2 \ge m_b^2$ two curves coincide
($\mu_b^2 = m_b^2$ and $\mu_b^2 = m_b^2/2$) since they are both
running with the same number of flavors ($n_f=5$) and they have the same
border condition. The curve using $\mu_b^2 = 2m_b^2$, however, still runs with
a smaller number of flavors ($n_f=4$) and so does not match the former two.
In the lower region $\mu_m^2 < m_b^2$ this is not happening, because even
if the number of flavors is now the same,
the border condition is specified above matching for $\mu_b^2 = m_b^2$ (in
$n_f=5$).
So, starting from $m_b^2$ and going downward, the central choice $\mu_b^2 =
m_b^2$ is matched first and then evolved, while the higher scale choice
$\mu_b^2 = 2m_b^2$ immediately runs with four light flavors at $m_b^2$. Thus
the difference consists just in the matching.

\section{Conclusions}
\label{sec:concl}
In this paper we presented a new \qcd{} tool to perform perturbative \dglap{}
evolution.
In \cref{sec:theory} we reviewed some theory aspects that are relevant
for this paper. In \cref{sec:pheno} we presented a few applications
of the implemented \eko{} features.

Most of the work done to develop \eko{} has been devoted to reproduce known
results from other programs (and slightly extending or amending them to have a
consistent behavior), in order to have a more flexible framework where to
implement new essential features for physics study (more on this in the Outlook
at the end of this section).
Benchmarks with already existing and widely used codes are shown in
\cref{sec:pheno:bench}, and demonstrated to be successful.
Further, the multiple options and configurations available are presented in
subsequent sections and discussed, all leading to known and understood
behaviors.

This does not mean that the current status of \eko{} does not expose any
novelty. \cref{tab:comp} summarizes a general comparison on specific features
between several evolution programs; we list only tools targeting the same scope
of \eko{}, that is unpolarized \pdf{} fitting.
It is exactly for this target (\pdf{} fitting) that \eko{} is optimized, and
among the others three specific features are outstanding: the solution in
$N$-space, the backward \vfns{} evolution, and the operator-oriented nature.

\eko{} is the first code to solve \dglap{} in Mellin space that has been
explicitly designed to be used for PDF fitting, and while this may seem
irrelevant, it has been explicitly picked as an improvement for \eko{} over the
similar codes.
There are multiple solutions that are only available in $x$-space by applying
numerical approximated procedures, making the exact solution the most reliable
one. In $N$-space this is not required, and the choice of the solution is
left completely up to the user, with no numerical deterioration among the
alternatives (as it was already for \pegasus{}), and thus it can be based on
theory considerations.
Moreover, the perturbative \qcd{} ingredients used in the evolution (like
anomalous dimensions) are often first computed in $N$-space, making them
available for \eko{} immediately, while a further complex transformation is
needed for usage in the other codes.

All the programs listed are able to preform backward evolution in \ffns{}, that
consists in swapping the integral evolution bounds, but the \vfns{} backward
evolution is a unique feature of \eko{}, which involves the non-trivial
inversion of the matching matrix, as outlined in \cref{sec:theory:matching}.

The reason why \eko{} is an operator-first framework is discussed in detail in
\cref{app:code:motiv}, but essentially it makes \eko{} particularly suited for
our target: repeated evaluation of evolution for \pdf{} fitting.
Producing only operators makes \eko{} less competitive for single one-shot
applications, but the optimal scaling with the size of the task (practically
constant, since the time consumed is dominated by the operator calculation)
makes it extremely good for massive evolution (and already good when evolving
$\order{10}$ elements).

It should be observed that while the choice of Python as programming language
particularly stands out among the other programs (all written in Fortran,
either 77 or 95), this is only a benefit from the point of view of project
management (being Python much expressive) and third parties contributions
(since its syntax is familiar to many).
Indeed, we make sure not to experience Python infamous performances, when it
comes to the most demanding tasks (like complex kernels evaluation, or Mellin
inverse integration) as they either use compiled extensions (e.g.\ those available
in \href{https://scipy.org/}{\texttt{scipy}}~\cite{2020SciPy-NMeth}) or they are compiled Just In
Time (JIT), using the \href{https://numba.pydata.org/}{\texttt{numba}}~\cite{numba} package.

\renewcommand{\thefootnote}{\alph{footnote}}
\begin{table}
    \centering
    \begin{tabular}{l|llllll}
    Feature & \eko{} & \apfel{}~\cite{Bertone:2013vaa} & \pegasus{}~\cite{Vogt:2004ns} & \hoppet{}~\cite{Salam:2008qg} & \qcdnum{}~\cite{Botje:2010ay} \\
    \hline
    input space & $x$ & $x$ & $N,x^{*}$ & $x$ & $x$ \\
    solution space & $N$ & $x$ & $N$ & $x$ & $x$ \\
    delivery space & $x$ & $x$ & $N,x$ & $x$ & $x$ \\
    delivery & $\vb E$ & $\vb f$\footnotemark[1] & $\vb {\tilde f},\vb f$ & $\vb f$\footnotemark[1] & $\vb f$ \\
    backward \ffns{} & \checkmark & \checkmark & \checkmark  & \checkmark & \checkmark \\
    backward \vfns{} & \checkmark & & & (\checkmark)\footnotemark[2] \\
    intrinsic evolution & \checkmark \\
    \hline
    prog. language & Python & Fortran 77 & Fortran 77 & Fortran 95 & Fortran 77\\
    produce \lhapdf{} grids & \checkmark & \checkmark  \\
    interpolation grids & \checkmark & \checkmark
    \end{tabular}
    \caption{Comparison between several evolution programs.
    The upper part of refers to some physical features: 
    by $x$ we mean the momentum fraction, $N$ the Mellin variables,
    $x^{*}$ denotes that \pegasus{} is able to deal with $x$-space input, 
    but only for fixed \pdf{} parametrization (see \cite{Vogt:2004ns}).
    $\vb E$ and $\vb f$ stands for evolution operators and \pdf{}s 
    respectively. 
    The lower part refers to program aspects, such as program language
    and interface with \lhapdf{}.
    }
    \label{tab:comp}
\end{table}
\footnotetext[1]{
    Both, \apfel{} and \hoppet{}, have an interface to access an evolution
    operator, but no one of the two can be used to store it and reuse it later on
    (this would require a dedicated interface).
}
\footnotetext[2]{
    \hoppet{} is able by default to do backward \vfns{}, but is not
    implementing intrinsic matching conditions (i.e.\ the contributions
    associated with the presence of an heavy flavor \pdf{}) nor the shifted
    matching scale.
}
\renewcommand{\thefootnote}{\arabic{footnote}}

While the main purpose of \eko{} is to evolve \pdf{}s, other \qcd{} ingredients
are required to perform the main task, like evolving the strong coupling
$\alpha_s$, running quark masses, or dealing with different flavor bases: they are
all provided to the end user, and for this reason actual results are shown in
this paper.

We remark that \eko{} is an open and living framework,
providing all ingredients as a public library, and accepting community
contributions, bug reports and feature requests, available through the public
\eko{} \href{https://github.com/N3PDF/eko/}{repository}.

\paragraph{Outlook} As outlined above \eko{} implements mostly well-known physics,
but we expect a series of upcoming project to build on the provided framework that will extend
the current knowledge on \pdf{}s.
Several features are already scheduled to be implemented,
and a few of them are already at an advanced stage:
the \nnnlo{} solution will be included as
soon as it becomes available~\cite{Moch:2021qrk}, while \nnnlo{} matching
conditions and strong coupling are already implemented and used in the
recent determination of the intrinsic charm content of the proton~\cite{Ball:2022qks}.

Another main goal of \eko{} is to provide a backbone in the determination of
\mhou{}, in the first place by allowing
the variation of the various scales used in the determination of evolved \pdf{}s,
that can be considered as an approximation to higher orders, implementing the
strategies described in \cite{AbdulKhalek:2019ihb}.
The variation of matching scales involved in \vfns{}
is already implemented and available.

Other planned features include: polarized evolution~\cite{Vogt:2008yw,Vogt:2014pha,Blumlein:2021ryt},
evolution of fragmentation functions~\cite{Mitov:2006ic,Moch:2007tx,Almasy:2011eq},
and the \qed$\otimes$\qcd{} evolution of
the photon-in-hadron distribution~\cite{Bertone:2017bme,Xie:2021equ,Cridge:2021pxm},
to estimate the impact of electro-weak corrections onto precision predictions.

\appendix
\section{Technical Overview}
\label{app:code}
An \eko{} is effectively a rank $5$ tensor $E_{\mu,i \alpha j \beta}$, that
evolves a PDF set from one given scale to a user specified list of final scales
$\mu$:
\begin{equation}
    f_{\mu,i\alpha} = E_{\mu,i \alpha j \beta} f^{(0)}_{j \beta}
\end{equation}
where $i$ and $j$ are indices on flavor, and $\alpha$ and $\beta$ are indices on the $x$-grid.

The computation of each rank $4$ tensor is almost independent:
In a \ffns{} for each target $\mu_F^2$ an operator $\vb{\tilde E}(\mu_{F,0}^2 \to \mu_F^2)$ is computed separately.
Instead, in a \vfns{} first a set of operators is computed, to evolve from the
initial scale to any matching scales (we call these \textit{threshold
operators}). Then, for each target $\mu_F^2$, an operator is computed
from the last intermediate matching scale to $\mu_F^2$; finally
they are composed together.

\subsection{Performance Motivations and Operator Specificity}
\label{app:code:motiv}

Before diving into the details of \eko{} performances there is a fundamental
point that has to be taken into account: \eko{} is somehow unique as an
evolution program, because its main and only output consists in evolution
\textit{operators}.

For this reason, a close comparison on performances with other evolution codes
(whose main purpose is the evolution of individual \pdf{}s) would be rather
unfair: evolving a single \pdf{} is comparable to the generation of the
transformation of a single direction in the \pdf{} space, while the operator acts
on the whole function space.

The motivation to primarily look for the operator itself relies on the specific
needs of a \pdf{} fit itself.
Indeed, a fit requires repeated usage of evolution for the $\chi^2$ evaluation
for each fit candidate, and a final evolution step for the generation of the
\pdf{} grids to deliver, as those used by \lhapdf{}~\cite{Buckley:2014ana}.
The first step has been automated long ago, by the generation of the
\fk{} tables (formerly done with \apfel{} evolution, through
\href{https://github.com/NNPDF/apfelcomb}{\apfelcomb}, inspired to
\cite{Bertone:2016lga}), that store \pdf{} evolution into the grids for
predictions, while the second was repeated at any fit, since for each fit is a
one-time operation (even though it is actually repeated for the number of Monte
Carlo replicas, or Hessian eigenvectors, whose typical sizes are reported in
\cref{tab:pdfmem}).

\begin{table}
    \centering
    \begin{tabular}{lr}
        \pdf{} set name & members \\
        \hline
        \texttt{CT18NNLO}~\cite{Hou:2019efy} & 59\\
        \texttt{MSHT20nnlo\_as118}~\cite{Bailey:2020ooq} & 65\\
        \texttt{NNPDF40\_nnlo\_as\_01180}~\cite{NNPDF:2021njg} & 101
    \end{tabular}
    \caption{Selected \pdf{} sets with their respective number of members}
    \label{tab:pdfmem}
\end{table}

Actually, both the operations of including evolution in theory grids and \pdf{}
grids generation can be further optimized, considering that the evolution only
depends on a small number of theory parameters, and so the operator does, such
that it can be generated only once, and then used over and over.

On top of replicas generation, the search towards an optimal fitting
methodology is an iterative process, that involves a large number number of fits.
Moreover, whatever program supports the generation of \fk{} tables has to
create some kind of evolution operator on its own (since the goal of \fk{}
tables is exactly to be \pdf{} agnostic).

So, the \eko{} work-flow is not a complete novelty, since it was preceded by
\apfel{} in-memory operator generation, but it is a further and strong
improvement in that direction: being operator-oriented from the beginning,
optimizations have been performed for this specific task\footnote{
E.g.\ internally integrating the minimal amount of anomalous dimensions required
for the operator determination, while still providing a flexible delivery on
all the output dimensions (re-interpolating the $x$ dependencies, or rotating
into different flavor bases).
}, and maintaining an
actual operator format, the operators reuse is possible even across the
boundary of \fk{} tables generation, and applied with benefit, e.g., for the
massive replicas set evolution (consider
\texttt{NNPDF40\_nnlo\_as\_01180\_1000}, that is a single set consisting of
1000 replicas, that can be evolved with a single operator instead of running
1000 times an evolution program, like all the other similar sets), but even
repeated fits.

While the benefit is limited for other use cases, any other highly iterative
phenomenological study, in which \pdf{} evolution is repeatedly evaluated from
different border conditions, would benefit from being backed by \eko{}, since
the cost of \dglap{} evaluation is paid only once (even though we are conscious
that this is mainly beneficial for \pdf{} fitting).

\subsection{Computation time}
\label{app:code:time}

As we said above the computation almost happens independently for each target $\mu_F^2$
and the amount of time required scales almost linearly with the number of requested
$\mu_F^2$, except for the thresholds operators in \vfns{} that are computed only once.

We call computing an operator with a fixed number of flavors \enquote{evolving
in a single patch}, since in a \vfns{} the evolution might span multiple patches.
When more than a single patch is involved, operators have to be joined at
matching scales $\mu_h$ with a non-trivial matching, that has to be
computed separately (these are part of the threshold operators).

Typical times required for these calculations in \eko{} are presented in
\cref{tab:time}.
As expected the complexity of the calculation grows with perturbative order,
and so even the time taken increases.
At \lo{} no matching conditions are needed, while for \nlo{} and
\nnlo{} they are computed once for each matching scale. 

\begin{table}[h]
    \centering
    \begin{tabular}{l|cc}
        & patch & matching \\
        \hline
        \lo{} & $\SI{10}{s}$ & $\varnothing$ \\
        \nlo{} & $\SI{45}{s}$ & $\SI{65}{s}$ \\
        \nnlo{} & $\SI{60}{s}$ & $\SI{75}{s}$ \\
    \end{tabular}
    \caption{Rough estimates of times taken by \eko{}, with an average sized
    $x$-grid of $50$ points and single core.}
    \label{tab:time}
\end{table}

We consider these time performances satisfactory, even though it is not
straightforward to compare \eko{} with the other evolution codes, as mentioned
in \cref{app:code:motiv}. As an example, \nnlo{}
evolution in $\mu_F^2 = \SI[parse-numbers=false]{1.65^2}{Gev^2} \to
\SI{100}{GeV^2}$ crossing the bottom matching at
$\SI[parse-numbers=false]{4.92^2}{Gev^2}$ takes $\sim \SI{60}{s} + \SI{135}{s}$
in \eko{} ($\SI{135}{s}$ for the thresholds operators initialization,
$\SI{60}{s}$ for the last patch). \apfel{} takes $\sim\SI{25}{s}$ on the same
custom interpolation $x$-grid (\apfel{} is able to perform significantly better
on a pre-defined, built-in grid).

This comparison shows that on the evolution of a single PDF \eko{} is not
really competitive, but the ratio is limited to $\sim 7.5$. 
However, we already pointed out that the two programs perform a rather
different task: computing a whole operator against a single \pdf{} evolution
(on which the benchmarking is done, only because both programs are able to
perform this simple task, but it is a worthless task for \eko{} usage).

The comparison is technically possible, but we do not encourage this kind of
benchmarks, because the typical task is actually different, and this motivates
the different performances.
\eko{} perform bad in the case of the single task, but with a perfect scaling
(negligible work needed for repeated evolution, practically constant), while
any other program would perform better for the atomic task, but with a linear
scaling in the number of objects to be evolved.

Each program should be selected having in mind the specific usage. \eko{} is
recommended for \pdf{} fitting, and repeated evolution in general.

The time measures in \cref{tab:time} and the rest of this
section have been obtained on a regular consumer laptop:
\begin{verbatim}
OS: Linux 5.13 Ubuntu 21.10 21.10 (Impish Indri)
CPU: (4) x64 Intel(R) Core(TM) i5-6267U CPU @ 2.90GHz
Memory: 7.56 GB
\end{verbatim}
None one of them is a careful benchmark, i.e.\ repeated multiple times, but is mainly
meant to show an order of magnitudes comparison.

\subsection{Memory footprint}
\label{app:code:memory}

Memory usage is dominated by the size of the final object produced,
since a much smaller internal representation is used during the computation.
The final object holds information about the rank $5$ operator, so it is
strictly dependent on the size of the interpolation $x$-grid and the amount of
target $\mu_F^2$ values.

For an average sized $x$-grid of $50$ points, and a single target $\mu_F^2$ the
size of the object in memory is of $\sim \SI{7.5}{MB}$, which scales linearly
with the amount of $\mu_F^2$ requested.
The dependency on the size of the $x$-grid is roughly quadratic.

\subsection{Storage}
\label{app:code:storage}

For permanent storage similar considerations applies with respect to the
memory object.
The main difference is that the object dumped by the \eko{} functions is always
compressed, leading to a size of $\sim \SI{500}{kB}$ for a single $\mu_F^2$,
which does not necessarily scales linearly with the amount $\mu_F^2$ since the
full rank $5$ tensor is compressed all-together (a linear scaling is just
the worst case).
Similar considerations applies to the dependency on the size of the $x$-grid.

\subsection{Possible Improvements}

There are a few easy directions to boost the current performances
of \eko{}, leveraging the $\mu_F^2$ splitting.

\paragraph{Jobs} To improve the speed of the computation, all the ingredients
of the final tensor (patches \& matching) can be computed by separate jobs, and
dispatched to different processors.
They just need to be joined at the very end in a simple linear algebra step.

Notice that the time measures presented in \cref{app:code:time} are obtained
with a fully sequential computation on a single processor, the only one
available at present time.

\paragraph{Memory} Since both the computation and the consumption of an \eko{}
can be done one $\mu_F^2$ at a time, it is possible to store each rank 4 tensor
on disk as soon as it is computed, and to load them in memory only while
applying them.\\

Both of these improvements are in the process of being implemented in \eko{}.

\section{User Manual}
\label{app:man}
In this section we provide an extremely brief manual about \eko{} usage.
We give here the instruction for the release version associated to this paper.
A more expanded and updated manual for the current version can be found in the on-line documentation:

\begin{center}
\url{https://eko.readthedocs.io/en/latest/overview/tutorials/index.html}
\end{center}

\subsection{Installation}

The installation provided to the user of the package is very simple.

We require only:
\begin{itemize}
    \item a working installation of Python 3\footnote{
            The exact Python version support can be found on the official
            \texttt{PyPI} (Python Package Index, the official Python registry)
            \eko{} page, see \url{https://pypi.org/project/eko/}.
        } and
    \item the official Python package manager
        \href{https://pip.pypa.io/}{\texttt{pip}}, usually bundled together any
        distribution of Python itself.
\end{itemize} 
You can check their availability on your system with:
\begin{lstlisting}[language=sh]
    python3 --version
    python3 -m pip --version
\end{lstlisting}
(for a non \texttt{sh} based environment, e.g.\ Windows, check the official
Python documentation).

The actual installation of the package can be obtained with the following
command:
\begin{lstlisting}[language=sh]
    python3 -m pip install eko==0.8.5
\end{lstlisting}

\subsection{Usage}

Using \eko{} as an evolution application is pretty simple, involving the call
of a single function to produce an evolution operator.

Nevertheless, it has been intentionally left up to the user the definition of
all the settings required for a run.
This might be confusing for a newcomer, but it is the best approach since the
actual settings might be different for any given application, and there is no
one that can be recommended as a \textit{best practice}; in order to not suggest
such an interpretation, we decided to not provide any defaults, and to require
the user to be aware of the whole set of settings.

We present now a minimal example, in which the settings are taken from the official
benchmarking setup in \cite{Giele:2002hx} (which is not worse nor better than any other choice).
To be able to access the toy PDF established in \cite{Giele:2002hx} you need to install in addition our benchmarking package:
\begin{lstlisting}[language=sh]
    python3 -m pip install banana-hep==0.6.6
\end{lstlisting}
Then you are set to run the following snippet:
\begin{lstlisting}[language=Python]
import numpy as np
import eko # of course we need eko
from banana import toy # load the official toy PDF

# theory setting
th_card = {
    "HQ": "POLE", # heavy quark mass scheme
    "IB": 0, # allow intrinsic bottom?
    "IC": 0, # allow intrinsic charm?
    "MaxNfAs": 6, # max number of flavor in alpha_s evolution
    "MaxNfPdf": 6, # max number of flavor in pdf evolution
    "ModEv": "EXA", # evolution mode
    "PTO": 0, # perturbation order of evolution: 0=LO, 1=NLO, 2=NNLO
    "Q0": np.sqrt(2.0), # fitting scale [GeV]
    "Qmb": 4.5, # MSbar reference scale for bottom mass [GeV]
    "Qmc": np.sqrt(2.0), # MSbar reference scale for charm mass [GeV]
    "Qmt": 175., # MSbar reference scale for top mass [GeV]
    "Qref": np.sqrt(2.0), # reference scale for alpha_s [GeV]
    "alphas": 0.35, # reference value for alpha_s
    "fact_to_ren_scale_ratio": 1.0, # scale variation ratio
    "kbThr": 1.0, # matching ratio to bottom mass
    "kcThr": 1.0, # matching ratio to charm mass
    "ktThr": 1.0, # matching ratio to top mass
    "mb": 4.5, # bottom mass [GeV]
    "mc": np.sqrt(2.0), # charm mass [GeV]
    "mt": 175., # top mass [GeV]
    "nf0": 3, # number of flavors at fitting scale
    "nfref": 3, # number of flavors at alpha_s reference scale
}
# operator settings
op_card = {
    "Q2grid": [10000.], # final scale grid
    "backward_inversion": "expanded", # backward inversion method
    "debug_skip_non_singlet": False, # debug option
    "debug_skip_singlet": False, # debug option
    "ev_op_iterations": 10, # number of iterations for solver
    "ev_op_max_order": 10, # expansion order of solver
    "interpolation_is_log": True, # use logarithmic interpolation?
    "interpolation_polynomial_degree": 4, # polynomial degree of interpolation
    "interpolation_xgrid": eko.interpolation.make_grid(30, 30), # interpolation grid ranging from 1e-7 to 1
}

# 1. compute the eko
evolution_operator = eko.run_dglap(th_card, op_card)
# 2. load the initial PDF (only for border condition) - any lhapdf like object will do
pdf = toy.mkPDF("", 0)
# 3. contract the given PDF with the eko
evolved_pdfs = evolution_operator.apply_pdf(pdf)
# then e.g. print the evolved gluon (pid = 21) at Q2 = 10000 GeV^2 for the first point x=1e-7
print(evolved_pdfs[10000.]["pdfs"][21][0])
# if we multiply with x again we obtain 1.3263e3
# The reference value from the official benchmark is 1.3272e3
# This is compatible with Fig. 1
print(1e-7 * evolved_pdfs[10000.]["pdfs"][21][0])
# or we can print the value of alpha_s at this scale, which also matches the reference value: 0.122306
print(evolution_operator["Q2grid"][10000.]["alphas"])
\end{lstlisting}

For more information about the settings, please refer to the online documentation.

Note, that the reference values for the gluon and the strong coupling are taken from \cite[Table 2]{Giele:2002hx}.

Also note, that the example may not work with newer version 
of the code, for which, instead, we recommend to follow the online tutorials.

\acknowledgments
\label{sec:ack}
We thank J.\ Cruz-Martinez for contributing to the development and S.\
Carrazza for suggesting the idea and providing valuable support. 
We thank S. Forte and J. Rojo for carefully proofreading the manuscript.
We acknowledge the NNPDF collaboration for valuable discussions
and comments.

F.~H. and A.~C. are supported by
the European Research Council under 
the European Union's Horizon 2020 research and innovation Programme
(grant agreement n.740006).
G.~M. is supported by NWO (Dutch Research Council).

\bibliographystyle{styles/JHEP}
\bibliography{bibliography/eko.bib,bibliography/refs.bib}

\providecommand{\href}[2]{#2}\begingroup\raggedright\begin{thebibliography}{10}

\bibitem{Gao:2017yyd}
J.~Gao, L.~Harland-Lang and J.~Rojo, \emph{{The Structure of the Proton in the
  LHC Precision Era}},
  \href{https://doi.org/10.1016/j.physrep.2018.03.002}{\emph{Phys. Rept.}
  {\bfseries 742} (2018) 1} [\href{https://arxiv.org/abs/1709.04922}{{\ttfamily
  1709.04922}}].

\bibitem{NNPDF:2017mvq}
{\scshape NNPDF} collaboration, \emph{{Parton distributions from high-precision
  collider data}},
  \href{https://doi.org/10.1140/epjc/s10052-017-5199-5}{\emph{Eur. Phys. J. C}
  {\bfseries 77} (2017) 663}
  [\href{https://arxiv.org/abs/1706.00428}{{\ttfamily 1706.00428}}].

\bibitem{Hou:2019efy}
T.-J.~Hou et~al., \emph{{New CTEQ global analysis of quantum chromodynamics
  with high-precision data from the LHC}},
  \href{https://doi.org/10.1103/PhysRevD.103.014013}{\emph{Phys. Rev. D}
  {\bfseries 103} (2021) 014013}
  [\href{https://arxiv.org/abs/1912.10053}{{\ttfamily 1912.10053}}].

\bibitem{Bailey:2020ooq}
S.~Bailey, T.~Cridge, L.A.~Harland-Lang, A.D.~Martin and R.S.~Thorne,
  \emph{{Parton distributions from LHC, HERA, Tevatron and fixed target data:
  MSHT20 PDFs}},
  \href{https://doi.org/10.1140/epjc/s10052-021-09057-0}{\emph{Eur. Phys. J. C}
  {\bfseries 81} (2021) 341}
  [\href{https://arxiv.org/abs/2012.04684}{{\ttfamily 2012.04684}}].

\bibitem{NNPDF:2021njg}
{\scshape NNPDF} collaboration, \emph{{The path to proton structure at 1\%
  accuracy}}, \href{https://doi.org/10.1140/epjc/s10052-022-10328-7}{\emph{Eur.
  Phys. J. C} {\bfseries 82} (2022) 428}
  [\href{https://arxiv.org/abs/2109.02653}{{\ttfamily 2109.02653}}].

\bibitem{AbdulKhalek:2019ihb}
{\scshape NNPDF} collaboration, \emph{{Parton Distributions with Theory
  Uncertainties: General Formalism and First Phenomenological Studies}},
  \href{https://doi.org/10.1140/epjc/s10052-019-7401-4}{\emph{Eur. Phys. J. C}
  {\bfseries 79} (2019) 931}
  [\href{https://arxiv.org/abs/1906.10698}{{\ttfamily 1906.10698}}].

\bibitem{Altarelli:1977zs}
G.~Altarelli and G.~Parisi, \emph{{Asymptotic Freedom in Parton Language}},
  \href{https://doi.org/10.1016/0550-3213(77)90384-4}{\emph{Nucl. Phys.}
  {\bfseries B126} (1977) 298}.

\bibitem{Gribov:1972ri}
V.N.~Gribov and L.N.~Lipatov, \emph{{Deep inelastic e p scattering in
  perturbation theory}}, {\emph{Sov. J. Nucl. Phys.} {\bfseries 15} (1972)
  438}.

\bibitem{Dokshitzer:1977sg}
Y.L.~Dokshitzer, \emph{{Calculation of the Structure Functions for Deep
  Inelastic Scattering and e+ e- Annihilation by Perturbation Theory in Quantum
  Chromodynamics.}}, {\emph{Sov. Phys. JETP} {\bfseries 46} (1977) 641}.

\bibitem{Ball:2008by}
{\scshape NNPDF} collaboration, \emph{{A Determination of parton distributions
  with faithful uncertainty estimation}},
  \href{https://doi.org/10.1016/j.nuclphysb.2008.09.037}{\emph{Nucl. Phys. B}
  {\bfseries 809} (2009) 1} [\href{https://arxiv.org/abs/0808.1231}{{\ttfamily
  0808.1231}}].

\bibitem{Ball:2010de}
R.D.~Ball, L.~Del~Debbio, S.~Forte, A.~Guffanti, J.I.~Latorre, J.~Rojo et~al.,
  \emph{{A first unbiased global NLO determination of parton distributions and
  their uncertainties}},
  \href{https://doi.org/10.1016/j.nuclphysb.2010.05.008}{\emph{Nucl. Phys. B}
  {\bfseries 838} (2010) 136}
  [\href{https://arxiv.org/abs/1002.4407}{{\ttfamily 1002.4407}}].

\bibitem{DelDebbio:2007ee}
{\scshape NNPDF} collaboration, \emph{{Neural network determination of parton
  distributions: The Nonsinglet case}},
  \href{https://doi.org/10.1088/1126-6708/2007/03/039}{\emph{JHEP} {\bfseries
  03} (2007) 039} [\href{https://arxiv.org/abs/hep-ph/0701127}{{\ttfamily
  hep-ph/0701127}}].

\bibitem{Vogt:2004mw}
A.~Vogt, S.~Moch and J.A.M.~Vermaseren, \emph{{The Three-loop splitting
  functions in QCD: The Singlet case}},
  \href{https://doi.org/10.1016/j.nuclphysb.2004.04.024}{\emph{Nucl. Phys.}
  {\bfseries B691} (2004) 129}
  [\href{https://arxiv.org/abs/hep-ph/0404111}{{\ttfamily hep-ph/0404111}}].

\bibitem{Moch:2004pa}
S.~Moch, J.A.M.~Vermaseren and A.~Vogt, \emph{{The Three loop splitting
  functions in QCD: The Nonsinglet case}},
  \href{https://doi.org/10.1016/j.nuclphysb.2004.03.030}{\emph{Nucl. Phys.}
  {\bfseries B688} (2004) 101}
  [\href{https://arxiv.org/abs/hep-ph/0403192}{{\ttfamily hep-ph/0403192}}].

\bibitem{Blumlein:2021enk}
J.~Bl\"umlein, P.~Marquard, C.~Schneider and K.~Sch\"onwald, \emph{{The
  three-loop unpolarized and polarized non-singlet anomalous dimensions from
  off shell operator matrix elements}},
  \href{https://doi.org/10.1016/j.nuclphysb.2021.115542}{\emph{Nucl. Phys. B}
  {\bfseries 971} (2021) 115542}
  [\href{https://arxiv.org/abs/2107.06267}{{\ttfamily 2107.06267}}].

\bibitem{Moch:2021qrk}
S.~Moch, B.~Ruijl, T.~Ueda, J.A.M.~Vermaseren and A.~Vogt, \emph{{Low moments
  of the four-loop splitting functions in QCD}},
  \href{https://doi.org/10.1016/j.physletb.2021.136853}{\emph{Phys. Lett. B}
  {\bfseries 825} (2022) 136853}
  [\href{https://arxiv.org/abs/2111.15561}{{\ttfamily 2111.15561}}].

\bibitem{Duhr:2021vwj}
C.~Duhr and B.~Mistlberger, \emph{{Lepton-pair production at hadron colliders
  at N$^{3}$LO in QCD}},
  \href{https://doi.org/10.1007/JHEP03(2022)116}{\emph{JHEP} {\bfseries 03}
  (2022) 116} [\href{https://arxiv.org/abs/2111.10379}{{\ttfamily
  2111.10379}}].

\bibitem{Ball:2022qks}
{\scshape NNPDF} collaboration, \emph{{Evidence for intrinsic charm quarks in
  the proton}}, \href{https://doi.org/10.1038/s41586-022-04998-2}{\emph{Nature}
  {\bfseries 608} (2022) 483}
  [\href{https://arxiv.org/abs/2208.08372}{{\ttfamily 2208.08372}}].

\bibitem{Carrazza_2020}
S.~Carrazza, E.R.~Nocera, C.~Schwan and M.~Zaro, \emph{Pineappl: combining ew
  and qcd corrections for fast evaluation of lhc processes},
  \href{https://doi.org/10.1007/jhep12(2020)108}{\emph{Journal of High Energy
  Physics} {\bfseries 2020} (2020) }.

\bibitem{christopher_schwan_2022_5846421}
C.~Schwan, A.~Candido, F.~Hekhorn and S.~Carrazza, \emph{N3pdf/pineappl:
  v0.5.0-beta.6},  Jan., 2022.
\newblock 10.5281/zenodo.5846421.

\bibitem{yadism}
A.~Candido et~al., ``{yadism: Yet Another DIS module}.'' {in preparation}.

\bibitem{Peskin:1995ev}
M.E.~Peskin and D.V.~Schroeder, \emph{{An Introduction to quantum field
  theory}}, Addison-Wesley, Reading, USA (1995).

\bibitem{Ellis:1996mzs}
R.K.~Ellis, W.J.~Stirling and B.R.~Webber, \emph{{QCD and collider physics}},
  vol.~8, Cambridge University Press (2, 2011),
  \href{https://doi.org/10.1017/CBO9780511628788}{10.1017/CBO9780511628788}.

\bibitem{Bertone:2013vaa}
V.~Bertone, S.~Carrazza and J.~Rojo, \emph{{APFEL: A PDF Evolution Library with
  QED corrections}},
  \href{https://doi.org/10.1016/j.cpc.2014.03.007}{\emph{Comput. Phys. Commun.}
  {\bfseries 185} (2014) 1647}
  [\href{https://arxiv.org/abs/1310.1394}{{\ttfamily 1310.1394}}].

\bibitem{Salam:2008qg}
G.P.~Salam and J.~Rojo, \emph{{A Higher Order Perturbative Parton Evolution
  Toolkit (HOPPET)}},
  \href{https://doi.org/10.1016/j.cpc.2008.08.010}{\emph{Comput. Phys. Commun.}
  {\bfseries 180} (2009) 120}
  [\href{https://arxiv.org/abs/0804.3755}{{\ttfamily 0804.3755}}].

\bibitem{Botje:2010ay}
M.~Botje, \emph{{QCDNUM: Fast QCD Evolution and Convolution}},
  \href{https://doi.org/10.1016/j.cpc.2010.10.020}{\emph{Comput.Phys.Commun.}
  {\bfseries 182} (2011) 490}
  [\href{https://arxiv.org/abs/1005.1481}{{\ttfamily 1005.1481}}].

\bibitem{Vogt:2004ns}
A.~Vogt, \emph{{Efficient evolution of unpolarized and polarized parton
  distributions with QCD-PEGASUS}},
  \href{https://doi.org/10.1016/j.cpc.2005.03.103}{\emph{Comput. Phys. Commun.}
  {\bfseries 170} (2005) 65}
  [\href{https://arxiv.org/abs/hep-ph/0408244}{{\ttfamily hep-ph/0408244}}].

\bibitem{Buckley:2011ms}
A.~Buckley et~al., \emph{{General-purpose event generators for LHC physics}},
  \href{https://doi.org/10.1016/j.physrep.2011.03.005}{\emph{Phys. Rept.}
  {\bfseries 504} (2011) 145}
  [\href{https://arxiv.org/abs/1101.2599}{{\ttfamily 1101.2599}}].

\bibitem{Carli:2010rw}
T.~Carli, D.~Clements, A.~Cooper-Sarkar, C.~Gwenlan, G.P.~Salam, F.~Siegert
  et~al., \emph{{A posteriori inclusion of parton density functions in NLO QCD
  final-state calculations at hadron colliders: The APPLGRID Project}},
  \href{https://doi.org/10.1140/epjc/s10052-010-1255-0}{\emph{Eur. Phys. J. C}
  {\bfseries 66} (2010) 503} [\href{https://arxiv.org/abs/0911.2985}{{\ttfamily
  0911.2985}}].

\bibitem{Britzger:2012bs}
{\scshape fastNLO} collaboration, \emph{{New features in version 2 of the
  fastNLO project}},  in \emph{{20th International Workshop on Deep-Inelastic
  Scattering and Related Subjects}}, pp.~217--221, 2012,
  \href{https://doi.org/10.3204/DESY-PROC-2012-02/165}{DOI}
  [\href{https://arxiv.org/abs/1208.3641}{{\ttfamily 1208.3641}}].

\bibitem{LagrangeInterpol}
W.~Edward, \emph{Vii. problems concerning interpolations},
  \href{https://doi.org/http://doi.org/10.1098/rstl.1779.0008}{\emph{Phil.
  Trans. R. Soc.} {\bfseries 69} (1779) 59–67}.

\bibitem{suli2003introduction}
E.~S{\"u}li and D.~Mayers, \emph{An Introduction to Numerical Analysis}, An
  Introduction to Numerical Analysis, Cambridge University Press (2003).

\bibitem{zbMATH02662492}
C.~Runge, \emph{{\"U}ber empirische {Funktionen} und die {Interpolation}
  zwischen {\"a}quidistanten {Ordinaten}.}, {\emph{Schl{\"o}milch Z.}
  {\bfseries 46} (1901) 224}.

\bibitem{Herzog:2017ohr}
F.~Herzog, B.~Ruijl, T.~Ueda, J.A.M.~Vermaseren and A.~Vogt, \emph{{The
  five-loop beta function of Yang-Mills theory with fermions}},
  \href{https://doi.org/10.1007/JHEP02(2017)090}{\emph{JHEP} {\bfseries 02}
  (2017) 090} [\href{https://arxiv.org/abs/1701.01404}{{\ttfamily
  1701.01404}}].

\bibitem{Luthe:2016ima}
T.~Luthe, A.~Maier, P.~Marquard and Y.~Schröder, \emph{{Towards the five-loop
  Beta function for a general gauge group}},
  \href{https://doi.org/10.1007/JHEP07(2016)127}{\emph{JHEP} {\bfseries 07}
  (2016) 127} [\href{https://arxiv.org/abs/1606.08662}{{\ttfamily
  1606.08662}}].

\bibitem{Baikov:2016tgj}
P.A.~Baikov, K.G.~Chetyrkin and J.H.~Kühn, \emph{{Five-Loop Running of the QCD
  coupling constant}},
  \href{https://doi.org/10.1103/PhysRevLett.118.082002}{\emph{Phys. Rev. Lett.}
  {\bfseries 118} (2017) 082002}
  [\href{https://arxiv.org/abs/1606.08659}{{\ttfamily 1606.08659}}].

\bibitem{Chetyrkin:2017bjc}
K.G.~Chetyrkin, G.~Falcioni, F.~Herzog and J.A.M.~Vermaseren, \emph{{Five-loop
  renormalisation of QCD in covariant gauges}},
  \href{https://doi.org/10.1007/JHEP10(2017)179}{\emph{JHEP} {\bfseries 10}
  (2017) 179} [\href{https://arxiv.org/abs/1709.08541}{{\ttfamily
  1709.08541}}].

\bibitem{Luthe:2017ttg}
T.~Luthe, A.~Maier, P.~Marquard and Y.~Schroder, \emph{{The five-loop Beta
  function for a general gauge group and anomalous dimensions beyond Feynman
  gauge}}, \href{https://doi.org/10.1007/JHEP10(2017)166}{\emph{JHEP}
  {\bfseries 10} (2017) 166}
  [\href{https://arxiv.org/abs/1709.07718}{{\ttfamily 1709.07718}}].

\bibitem{Buza_1998}
M.~Buza, Y.~Matiounine, J.~Smith and W.L.~van Neerven, \emph{Charm
  electroproduction viewed in the variable-flavour number scheme versus
  fixed-order perturbation theory},
  \href{https://doi.org/10.1007/bf01245820}{\emph{The European Physical Journal
  C} {\bfseries 1} (1998) 301–320}.

\bibitem{Ball:2016neh}
{\scshape NNPDF} collaboration, \emph{{A Determination of the Charm Content of
  the Proton}},
  \href{https://doi.org/10.1140/epjc/s10052-016-4469-y}{\emph{Eur. Phys. J. C}
  {\bfseries 76} (2016) 647}
  [\href{https://arxiv.org/abs/1605.06515}{{\ttfamily 1605.06515}}].

\bibitem{Alekhin:2010sv}
S.~Alekhin and S.~Moch, \emph{{Heavy-quark deep-inelastic scattering with a
  running mass}},
  \href{https://doi.org/10.1016/j.physletb.2011.04.026}{\emph{Phys. Lett. B}
  {\bfseries 699} (2011) 345}
  [\href{https://arxiv.org/abs/1011.5790}{{\ttfamily 1011.5790}}].

\bibitem{Vermaseren:1997fq}
J.A.M.~Vermaseren, S.A.~Larin and T.~van Ritbergen, \emph{{The four loop quark
  mass anomalous dimension and the invariant quark mass}},
  \href{https://doi.org/10.1016/S0370-2693(97)00660-6}{\emph{Phys. Lett. B}
  {\bfseries 405} (1997) 327}
  [\href{https://arxiv.org/abs/hep-ph/9703284}{{\ttfamily hep-ph/9703284}}].

\bibitem{Schroder:2005hy}
Y.~Schroder and M.~Steinhauser, \emph{{Four-loop decoupling relations for the
  strong coupling}},
  \href{https://doi.org/10.1088/1126-6708/2006/01/051}{\emph{JHEP} {\bfseries
  01} (2006) 051} [\href{https://arxiv.org/abs/hep-ph/0512058}{{\ttfamily
  hep-ph/0512058}}].

\bibitem{Chetyrkin:2005ia}
K.~Chetyrkin, J.H.~Kuhn and C.~Sturm, \emph{{QCD decoupling at four loops}},
  \href{https://doi.org/10.1016/j.nuclphysb.2006.03.020}{\emph{Nucl. Phys. B}
  {\bfseries 744} (2006) 121}
  [\href{https://arxiv.org/abs/hep-ph/0512060}{{\ttfamily hep-ph/0512060}}].

\bibitem{Giele:2002hx}
W.~Giele et~al., \emph{{The QCD / SM working group: Summary report}},  in
  \emph{{2nd Les Houches Workshop on Physics at TeV Colliders}}, pp.~275--426,
  4, 2002 [\href{https://arxiv.org/abs/hep-ph/0204316}{{\ttfamily
  hep-ph/0204316}}].

\bibitem{Dittmar:2005ed}
M.~Dittmar et~al., \emph{{Working Group I: Parton distributions: Summary report
  for the HERA LHC Workshop Proceedings}}, {\emph{{WGI}} (2005) }
  [\href{https://arxiv.org/abs/hep-ph/0511119}{{\ttfamily hep-ph/0511119}}].

\bibitem{Diehl:2021gvs}
M.~Diehl, R.~Nagar and F.J.~Tackmann, \emph{{ChiliPDF: Chebyshev interpolation
  for parton distributions}},
  \href{https://doi.org/10.1140/epjc/s10052-022-10223-1}{\emph{Eur. Phys. J. C}
  {\bfseries 82} (2022) 257}
  [\href{https://arxiv.org/abs/2112.09703}{{\ttfamily 2112.09703}}].

\bibitem{Buckley:2014ana}
A.~Buckley, J.~Ferrando, S.~Lloyd, K.~Nordstr\"om, B.~Page, M.~R\"ufenacht
  et~al., \emph{{LHAPDF6: parton density access in the LHC precision era}},
  \href{https://doi.org/10.1140/epjc/s10052-015-3318-8}{\emph{Eur. Phys. J. C}
  {\bfseries 75} (2015) 132} [\href{https://arxiv.org/abs/1412.7420}{{\ttfamily
  1412.7420}}].

\bibitem{Bonvini:2012sh}
M.~Bonvini, \emph{{Resummation of soft and hard gluon radiation in perturbative
  QCD}}, Ph.D. thesis, Genoa U., 2012.
\newblock \href{https://arxiv.org/abs/1212.0480}{{\ttfamily 1212.0480}}.

\bibitem{Bertone:2014zva}
V.~Bertone, R.~Frederix, S.~Frixione, J.~Rojo and M.~Sutton, \emph{{aMCfast:
  automation of fast NLO computations for PDF fits}},
  \href{https://doi.org/10.1007/JHEP08(2014)166}{\emph{JHEP} {\bfseries 08}
  (2014) 166} [\href{https://arxiv.org/abs/1406.7693}{{\ttfamily 1406.7693}}].

\bibitem{Bierenbaum:2009zt}
I.~Bierenbaum, J.~Blumlein and S.~Klein, \emph{{The Gluonic Operator Matrix
  Elements at O(alpha(s)**2) for DIS Heavy Flavor Production}},
  \href{https://doi.org/10.1016/j.physletb.2009.01.057}{\emph{Phys. Lett. B}
  {\bfseries 672} (2009) 401}
  [\href{https://arxiv.org/abs/0901.0669}{{\ttfamily 0901.0669}}].

\bibitem{Bierenbaum:2009mv}
I.~Bierenbaum, J.~Blumlein and S.~Klein, \emph{{Mellin Moments of the
  O(alpha**3(s)) Heavy Flavor Contributions to unpolarized Deep-Inelastic
  Scattering at Q**2 \ensuremath{>}\ensuremath{>} m**2 and Anomalous
  Dimensions}},
  \href{https://doi.org/10.1016/j.nuclphysb.2009.06.005}{\emph{Nucl. Phys. B}
  {\bfseries 820} (2009) 417}
  [\href{https://arxiv.org/abs/0904.3563}{{\ttfamily 0904.3563}}].

\bibitem{Ablinger:2010ty}
J.~Ablinger, J.~Blumlein, S.~Klein, C.~Schneider and F.~Wissbrock, \emph{{The
  $O(\alpha_s^3)$ Massive Operator Matrix Elements of $O(n_f)$ for the
  Structure Function $F_2(x,Q^2)$ and Transversity}},
  \href{https://doi.org/10.1016/j.nuclphysb.2010.10.021}{\emph{Nucl. Phys. B}
  {\bfseries 844} (2011) 26} [\href{https://arxiv.org/abs/1008.3347}{{\ttfamily
  1008.3347}}].

\bibitem{Ablinger:2014vwa}
J.~Ablinger, A.~Behring, J.~Bl\"umlein, A.~De~Freitas, A.~Hasselhuhn, A.~von
  Manteuffel et~al., \emph{{The 3-Loop Non-Singlet Heavy Flavor Contributions
  and Anomalous Dimensions for the Structure Function $F_2(x,Q^2)$ and
  Transversity}},
  \href{https://doi.org/10.1016/j.nuclphysb.2014.07.010}{\emph{Nucl. Phys. B}
  {\bfseries 886} (2014) 733}
  [\href{https://arxiv.org/abs/1406.4654}{{\ttfamily 1406.4654}}].

\bibitem{Ablinger:2014uka}
J.~Ablinger, J.~Bl\"umlein, A.~De~Freitas, A.~Hasselhuhn, A.~von Manteuffel,
  M.~Round et~al., \emph{{The $O(\alpha_s^3 T_F^2)$ Contributions to the
  Gluonic Operator Matrix Element}},
  \href{https://doi.org/10.1016/j.nuclphysb.2014.05.028}{\emph{Nucl. Phys. B}
  {\bfseries 885} (2014) 280}
  [\href{https://arxiv.org/abs/1405.4259}{{\ttfamily 1405.4259}}].

\bibitem{Behring:2014eya}
A.~Behring, I.~Bierenbaum, J.~Bl\"umlein, A.~De~Freitas, S.~Klein and
  F.~Wi\ss{}brock, \emph{{The logarithmic contributions to the $O(\alpha^3_s)$
  asymptotic massive Wilson coefficients and operator matrix elements in deeply
  inelastic scattering}},
  \href{https://doi.org/10.1140/epjc/s10052-014-3033-x}{\emph{Eur. Phys. J. C}
  {\bfseries 74} (2014) 3033}
  [\href{https://arxiv.org/abs/1403.6356}{{\ttfamily 1403.6356}}].

\bibitem{Ablinger_2014}
J.~Ablinger, J.~Blümlein, A.~De~Freitas, A.~Hasselhuhn, A.~von Manteuffel,
  M.~Round et~al., \emph{The transition matrix element $a_{gq}(n)$ of the
  variable flavor number scheme at $o(\alpha_s^3)$},
  \href{https://doi.org/10.1016/j.nuclphysb.2014.02.007}{\emph{Nuclear Physics
  B} {\bfseries 882} (2014) 263–288}.

\bibitem{Ablinger_2015}
J.~Ablinger, A.~Behring, J.~Blümlein, A.~De~Freitas, A.~von Manteuffel and
  C.~Schneider, \emph{The 3-loop pure singlet heavy flavor contributions to the
  structure function f2(x,q2) and the anomalous dimension},
  \href{https://doi.org/10.1016/j.nuclphysb.2014.10.008}{\emph{Nuclear Physics
  B} {\bfseries 890} (2015) 48–151}.

\bibitem{Blumlein:2017wxd}
J.~Bl\"umlein, J.~Ablinger, A.~Behring, A.~De~Freitas, A.~von Manteuffel,
  C.~Schneider et~al., \emph{{Heavy Flavor Wilson Coefficients in
  Deep-Inelastic Scattering: Recent Results}},
  \href{https://doi.org/10.22323/1.308.0031}{\emph{PoS} {\bfseries QCDEV2017}
  (2017) 031} [\href{https://arxiv.org/abs/1711.07957}{{\ttfamily
  1711.07957}}].

\bibitem{LHCHiggsCrossSectionWorkingGroup:2016ypw}
{\scshape LHC Higgs Cross Section Working Group} collaboration, \emph{{Handbook
  of LHC Higgs Cross Sections: 4. Deciphering the Nature of the Higgs Sector}},
   \href{https://arxiv.org/abs/1610.07922}{{\ttfamily 1610.07922}}.

\bibitem{2020SciPy-NMeth}
P.~Virtanen, R.~Gommers, T.E.~Oliphant, M.~Haberland, T.~Reddy, D.~Cournapeau
  et~al., \emph{{{SciPy} 1.0: Fundamental Algorithms for Scientific Computing
  in Python}}, \href{https://doi.org/10.1038/s41592-019-0686-2}{\emph{Nature
  Methods} {\bfseries 17} (2020) 261}.

\bibitem{numba}
S.K.~Lam, A.~Pitrou and S.~Seibert, \emph{Numba: A llvm-based python jit
  compiler},  in \emph{Proceedings of the Second Workshop on the LLVM Compiler
  Infrastructure in HPC}, LLVM '15, (New York, NY, USA), Association for
  Computing Machinery, 2015,
  \href{https://doi.org/10.1145/2833157.2833162}{DOI}.

\bibitem{Vogt:2008yw}
A.~Vogt, S.~Moch, M.~Rogal and J.A.M.~Vermaseren, \emph{{Towards the NNLO
  evolution of polarised parton distributions}},
  \href{https://doi.org/10.1016/j.nuclphysbps.2008.09.097}{\emph{Nucl. Phys. B
  Proc. Suppl.} {\bfseries 183} (2008) 155}
  [\href{https://arxiv.org/abs/0807.1238}{{\ttfamily 0807.1238}}].

\bibitem{Vogt:2014pha}
A.~Vogt, S.~Moch and J.A.M.~Vermaseren, \emph{{A calculation of the three-loop
  helicity-dependent splitting functions in QCD}},
  \href{https://doi.org/10.22323/1.211.0040}{\emph{PoS} {\bfseries LL2014}
  (2014) 040} [\href{https://arxiv.org/abs/1405.3407}{{\ttfamily 1405.3407}}].

\bibitem{Blumlein:2021ryt}
J.~Bl\"umlein, P.~Marquard, C.~Schneider and K.~Sch\"onwald, \emph{{The
  three-loop polarized singlet anomalous dimensions from off-shell operator
  matrix elements}}, \href{https://doi.org/10.1007/JHEP01(2022)193}{\emph{JHEP}
  {\bfseries 01} (2022) 193}
  [\href{https://arxiv.org/abs/2111.12401}{{\ttfamily 2111.12401}}].

\bibitem{Mitov:2006ic}
A.~Mitov, S.~Moch and A.~Vogt, \emph{{Next-to-Next-to-Leading Order Evolution
  of Non-Singlet Fragmentation Functions}},
  \href{https://doi.org/10.1016/j.physletb.2006.05.005}{\emph{Phys. Lett. B}
  {\bfseries 638} (2006) 61}
  [\href{https://arxiv.org/abs/hep-ph/0604053}{{\ttfamily hep-ph/0604053}}].

\bibitem{Moch:2007tx}
S.~Moch and A.~Vogt, \emph{{On third-order timelike splitting functions and
  top-mediated Higgs decay into hadrons}},
  \href{https://doi.org/10.1016/j.physletb.2007.10.069}{\emph{Phys. Lett. B}
  {\bfseries 659} (2008) 290}
  [\href{https://arxiv.org/abs/0709.3899}{{\ttfamily 0709.3899}}].

\bibitem{Almasy:2011eq}
A.A.~Almasy, S.~Moch and A.~Vogt, \emph{{On the Next-to-Next-to-Leading Order
  Evolution of Flavour-Singlet Fragmentation Functions}},
  \href{https://doi.org/10.1016/j.nuclphysb.2011.08.028}{\emph{Nucl. Phys. B}
  {\bfseries 854} (2012) 133}
  [\href{https://arxiv.org/abs/1107.2263}{{\ttfamily 1107.2263}}].

\bibitem{Bertone:2017bme}
{\scshape NNPDF} collaboration, \emph{{Illuminating the photon content of the
  proton within a global PDF analysis}},
  \href{https://doi.org/10.21468/SciPostPhys.5.1.008}{\emph{SciPost Phys.}
  {\bfseries 5} (2018) 008} [\href{https://arxiv.org/abs/1712.07053}{{\ttfamily
  1712.07053}}].

\bibitem{Xie:2021equ}
{\scshape CTEQ-TEA} collaboration, \emph{{Photon PDF within the CT18 global
  analysis}}, \href{https://doi.org/10.1103/PhysRevD.105.054006}{\emph{Phys.
  Rev. D} {\bfseries 105} (2022) 054006}
  [\href{https://arxiv.org/abs/2106.10299}{{\ttfamily 2106.10299}}].

\bibitem{Cridge:2021pxm}
T.~Cridge, L.A.~Harland-Lang, A.D.~Martin and R.S.~Thorne, \emph{{QED parton
  distribution functions in the MSHT20 fit}},
  \href{https://doi.org/10.1140/epjc/s10052-022-10028-2}{\emph{Eur. Phys. J. C}
  {\bfseries 82} (2022) 90} [\href{https://arxiv.org/abs/2111.05357}{{\ttfamily
  2111.05357}}].

\bibitem{Bertone:2016lga}
V.~Bertone, S.~Carrazza and N.P.~Hartland, \emph{{APFELgrid: a high performance
  tool for parton density determinations}},
  \href{https://doi.org/10.1016/j.cpc.2016.10.006}{\emph{Comput. Phys. Commun.}
  {\bfseries 212} (2017) 205}
  [\href{https://arxiv.org/abs/1605.02070}{{\ttfamily 1605.02070}}].

\end{thebibliography}\endgroup

\listoffixmes

\end{document}